\documentclass[a4paper,11pt,pointlessnumbers]{scrartcl}
\usepackage{amsmath}
\usepackage{amsfonts,txfonts}
\usepackage{amssymb,wasysym}
\usepackage{t1enc}
\usepackage[latin1]{inputenc}
\usepackage[english]{babel}
\usepackage[numbers]{natbib}
\usepackage{amssymb,stmaryrd}

\usepackage[final]{pdfpages}
\usepackage{caption}
\usepackage{url}

\usepackage[colorlinks=false, linktocpage=true,backref,pagebackref,hidelinks]{hyperref}

\newcommand{\beq}{\begin{equation}}
\newcommand{\eeq}{\end{equation}}
\newcommand{\beqarr}{\begin{eqnarray}}
\newcommand{\eeqarr}{\end{eqnarray}}
\newcommand{\beqa}{\begin{eqnarray*}}
\newcommand{\eeqa}{\end{eqnarray*}}
\begin{document}
\thispagestyle{empty}

\title{\bf \large  A scalar field inducing a non-metrical contribution to  gravitational acceleration and a compatible add-on to light deflection}
\author{\normalsize Erhard Scholz\footnote{University of Wuppertal, Faculty of  Math./Natural Sciences, and Interdisciplinary Centre for History and Philosophy of Science, \quad  scholz@math.uni-wuppertal.de}}

\date{\small 20. 03. 2020 }
\maketitle

\begin{abstract}
{\small 
A scalar field model for explaining the anomalous acceleration  and light deflection at galactic and cluster scales, without further dark matter,  is presented. It is formulated in a scale covariant scalar tensor theory of gravity in the framework of integrable Weyl geometry and presupposes two different phases for the scalar field, like the superfluid approach of Berezhiani/Khoury.  In   low acceleration regimes of static gravitational fields (in the Einstein frame) with accordingly low values of the scalar field gradient, the scalar field Lagrangian combines  a cubic kinetic term  similar to the  ``a-quadratic''  Lagrangian  used in the  first covariant generalization of MOND (RAQUAL)  \citep{Bekenstein/Milgrom:1984} and  a second order derivative term introduced by  Novello et al. in the context of a Weyl geometric approach to cosmology \citep{Novello/Oliveira_ea:1993,Oliveira/Salim/Sautu:1997}.
In varying with regard to $\phi$ the latter is variationally equivalent to a  first order expression. The  scalar field equation thus remains of order two. In the  Einstein frame it assumes the form of a covariant generalization of the Milgrom equation known from the classical MOND approach in the deep MOND regime. It implies a corresponding ``non-metrical'' contribution to the acceleration of free fall trajectories. In contrast to pure RAQUAL, the second order derivative term of the Lagrangian leads to a non-negligible contribution to the energy momentum tensor and an {\em add-on to the light deflection potential} in beautiful agreement with the dynamics of low velocity trajectories. Although the model takes up important ingredients from the usual RAQUAL approach, it  differs essentially from the latter. -- 
In  higher sectional curvature regions, respectively for   higher accelerations in static fields, the scalar field Lagrangian consists  of  a Jordan-Brans-Dicke  term  with sufficiently high value of the  JBD-constant  to satisfy empirical constraints. Here the dynamics  agrees   effectively  with the one of  Einstein gravity. 
}
\end{abstract}

\setcounter{tocdepth}{2}
{\small \tableofcontents
\noindent
}

\section*{Introduction}
\addcontentsline{toc}{section}{\protect\numberline{}Introduction}

\vspace*{0.5em}
A  cubic term in the partial derivatives of a scalar field $\phi$ was  introduced by M. Milgrom and J. Bekenstein in their Lagrangian formulation of  Milgrom's ``modified Newtonian dynamics'' (MOND) and its first relativistic generalization RAQUAL  (an acronym for a ``relativistic a-quadratic Lagrangian'')   \citep{Bekenstein/Milgrom:1984}. Similar cubic expressions in first derivatives of additional fields have become a generic feature of models  attempting to  explain  the flat rotation curves of galaxies by a modification of gravity rather than by  dark matter. The physical origin of these Lagrangians and the reason for the different types of dynamics in extremely weak gravitational regimes (MOND-like) and in regions with stronger gravitation  (Newtonian or Einsteinian) have remained  open.   J. Khoury, L Berezhiani et al.  propose an interesting path for shedding light on both questions \citep{Berezhiani/Khoury:2015,Berezhiani/Khoury:2016,Berezhiani/Famaey/Khoury:2017,Berezhiani/Khoury:2019}. According to them,
   fractional powers of $(\partial \phi) ^2$ are not uncommon in the effective Lagrangians of dynamical excitations in superfluids; moreover the formation of a superfluid phase of a (hypothetical) underlying substrate is dependent on special conditions which make it natural  to assume  different phases for the  substrate (which they still leave open) in gravitationally strongly differing spacetime regions.
  They also try to connect their superfluid  dark matter with  modified gravity of a MONDian type (in their case TeVeS).  
  
The authors of RAQUAL themselves saw and criticized  the limitations of their   approach. Although it leads to a MOND-like modification  (of Newton/Einstein gravity) for low velocity orbits, it is unable to
explain the observed additional gravitational lensing effects.  Moreover it admits   excitations with superluminal propagation. Various other  relativistic generalizations of MOND, in particular  TeVeS (``Tensor-vector-scalar'') theory \citep{Bekenstein:2004,Skordis:2008},   ``Einstein-aether'' theories \citep{Jacobson/Mattingley:2001}, have been proposed to cure these deficiencies.   These models often introduce an additional vector field  besides one or more scalar ones as part of the gravitational structure, and use strange geometries as their spacetime framework with, e.g.,  two different metrics  with a non-conformal transformation  between them (TeVeS), 
 or a breaking of Lorentz invariance in the infinitesimal domain (Einstein ether).  Motivation for some features  of these generalizations came from cosmology rather than from (or in addition to) astrophysics \citep[sec. 7]{Famaey/McGaugh:MOND}, \citep[sec. 3]{Clifton/Ferreira_ea:2012}. 

The present paper comes back to presupposing only 
 one additional scalar field like RAQUAL, but with a a different Lagrangian density and assuming two different phases for the scalar field, like in the superfluid approach of Berezhiani/Khoury.  In the Einstein frame and very low acceleration regions of static gravitational fields (accordingly for very low values of the sectional curvature for the Einstein metric), the scalar field Lagrangian combines  a cubic kinetic term    $(\partial \phi)^3$, similar to RAQUAL in  the deep MOND regime,  with a  second order derivative self-interaction term (type $\partial^2 \phi$) first studied by   Novello et al. in the context of a Weyl geometric approach to cosmology \citep{Novello/Oliveira_ea:1993,Oliveira/Salim/Sautu:1997},  and a conformally coupled second order term. 
This phase  will be called the {\em MG} (modified gravity) {\em regime} of the present model. 
  All terms of the  Lagrangian are formulated  in a locally scale covariant form  with integrable Weyl geometry as the geometrical framework for the spacetime structure and scale covariant derivatives of the fields.  In  higher sectional curvature regions, respectively for   higher accelerations in static fields, the scalar field Lagrangian consists  of  a Jordan-Brans-Dicke (JBD) term  with sufficiently high value of the   coefficient   to satisfy empirical constraints which lead effectively to the dynamics of Einstein gravity. This will be called the {\em eEG} (effective Einstein gravity)  {\em regime}.
  
  The scalar field,  $ \phi= C\, e^{-\sigma}$ in Riemann gauge (Jordan frame), plays a crucial role in modifying the spacetime geometry. Its logarithm $\sigma$  is the potential of an integrable Weylian scale connection $\varphi= - d \sigma$ in Einstein frame/gauge. In the eEG-regime its effects are negligible, while in the MG  regime it modifies the free fall trajectories in Einstein gauge (Einstein frame) similar to RAQUAL. While varying with regard to $\phi$   the  second order scale covariant  derivative $D_{\nu}D^{\nu}\phi$ 
 is variationally equivalent to a  first order expression, because the second order partial derivatives $\partial_{\nu}\partial^{\nu}\phi$ occur in a divergence term involving the metric $g$. The  scalar field equation thus remains of order two.  This was already observed by the Brazilian group of physicists and used for constructing  their cosmological model.   
 
  Due to the combined cubic and second order Lagrangian (and adding the conformally coupled second order term), the scalar field equation takes on  the form of a covariant generalization of the Milgrom equation, if one goes over to the Einstein frame. Its flat (Euclidean) counter-part is known from the deep MOND regime in the classical MOND approach and  serves as an approximation for regions in which the Riemannian curvature components are  small enough for  allowing  a Newton approximation of the Einstein equation. Both together will be called a {\em Newton-Milgrom approximation} of the present model in the MG-regime. The solution of the Milgrom equation  implies a corresponding non-metrical contribution to the acceleration of free fall trajectories.
  
In contrast to pure RAQUAL, the second order ($\partial^2\phi$) term leads to a non-negligible contribution to the energy momentum tensor of the scalar field. It comes as a surprise that, although the scalar field energy momentum does not contribute to the  source of the Newton approximation  ($\rho + \sum_1^3 p_j$ of the scalar field vanishes), it does so for the calculation of the gravitational light deflection in the MG-regime.   In the central symmetric case it leads to an {\em add-on of the light deflection potential} in beautiful agreement with the dynamics of low velocity trajectories. We may  conjecture that this also holds generally.  
   
Finally the scalar field energy seems to  play an important  role for the  modification of gravity also in hierarchical structured systems. In  very weak acceleration configurations the solutions of the Milgrom approximations at different levels, e.g., galactic and cluster, may (tentatively) be  superimposed linearly. For clusters this leads to an estimate of the overall modification of gravity different from other MOND-like theories, measurable in a larger ``total mass'' (only partially phantom) attributed to the gravitational effects  of the scalar field via its influence on the non-metrical component of the affine connection. 
The energy-momentum of the scalar field can be considered as the expression of a kind of ``dark matter'' {\em sui generis} with repercussion in particular for cluster dynamics. 
All in all the present model differs considerably from RAQUAL although it  takes up important ingredients from the  the latter.

The organization of the paper is as follows: 
The basic features of the present approach are presented in the first section. Sec. \ref{subsection Framework}, supported by  appendix \ref{Appendix IWG}, gives a refreshment of the geometrical framework of integrable Weyl geometry and introduces the respective notations. In a scale gauge (roughly corresponding the choice of a {\em frame} in Jordan-Brans-Dicke theory) the complete data of a Weylian metric are given in the integrable case by a pair $(g, \varphi)$ consisting  of a (pseudo-) Riemannian metric $g$ and a closed differential 1-form, or in physics terminology a  pure gauge co-vector field $\varphi$. 
Sec. \ref{subsection Lagrangian} formulates the  scale invariant Lagrangian densities  of the  model, in particular those for the  
two different phases of the scalar field. The  derivation of the modified (Weyl geometric) Einstein equation and the scalar field equation for both regimes follows in secs. \ref{subsection Einstein equation}, \ref{subsection scalar field equations}.
In the first  (eEG) regime the dynamics is like in JBD theory; we therefore have to assume a sufficiently high value for the coupling coefficient of the JBD kinetic term in order to arrive effectively at the dynamics of Einstein gravity. In the second (MG) regime we derive a covariant generalization, eq. (\ref{eq covariant Milgrom equation}), of the non-linear Poisson equation known from classical MOND kinematics. By obvious reasons it will be called the {\em covariant Milgrom equation}.  Free fall trajectories in the present (integrable Weyl geometric) framework are studied in sec. \ref{subsection trajectories}. In the Einstein gauge (comparable to the Einstein frame of JBD theory) an additional term of the acceleration of test bodies arises. It is due to the scale connection determined by the scalar field, eq. (\ref{eq add acc 1}). Finally a discussion of the question in which scale  measurable quantities  are most directly expressed leads us again to the Einstein gauge (sec. \ref{subsection Einstein gauge}).

The second section deals with the dynamics in the MG regime. Sec. \ref{subsection MG} analyses  the  modification due to the scalar field
of the Einstein equation for the  Riemannian part  $g$ of the Weylian metric. In sec. \ref{subsection Newton-Milgrom} we turn towards  the quasi-static weak field approximations for the Einstein equation and the Milgrom equation. The first one boils down to a  Newton approximation like in Einstein  gravity. The second one has the form of the non-linear Poisson equation of deep MOND.   Both together constitute the {\em Newton-Milgrom approximation} of our approach. The solution of the Milgrom approximation leads to an additional acceleration of  a form known from the deep MOND regime in the usual MOND approach.
The energy momentum tensor of the scalar field in the Milgrom approximation has intriguing properties  (sec. \ref{subsection enhanced momentum}f.). Although at a first glance its energy density might seem consistent with the additional acceleration (it agrees with the phantom energy density associated to the additional acceleration of the scale connection), it does not contribute to the source term of the Newton approximation because of its negative pressure terms.
 On the other hand, it does contribute to light deflection.
 An investigation of the spatial components of Riemannian component of the metric in the weak field approximation and Einstein gauge for the central symmetric case leads to a light deflection potential consistent with the additional acceleration of the scale connection (sec. \ref{subsection light deflection}).  

This is a peculiar result for the present model, up to now unexpected for  relativistic generalizations of MOND with only one additional scalar field. The respective MOND-typical interpolation functions and their feasibility for galactic dynamics are discussed in section  \ref{subsection MOND} and \ref{subsection galaxies}. It also may have important repercussions for the dynamics of galaxy clusters (secs. \ref{subsection clusters-1}, \ref{subsection clusters-2}).  Additional questions not dealt with in detail in this paper are shortly discussed in the final  section \ref{section Discussion}.



\section{\small The approach \label{section Approach}}

\subsection{\small Framework \label{subsection Framework}}
We use a scale covariant generalization of Einstein gravity formulated in the framework of  integrable Weyl geometry  with one additional degree of freedom only, incorporated by a gravitationally coupled scalar field  (see appendix \ref{Appendix IWG}).  The Weylian metric is given in any (scale) gauge  by the data $(g, \varphi)$,  
 where $g=  g_{\mu \nu}dx^{\mu}dx^{\nu}$ is the {\em Riemannian component} of the (Weylian) metric, here of Lorentzian signature type $(-+++)$, and  $\varphi$  represents an integrable Weylian {\em scale connection}, given by a closed differential 1-form (a pure gauge co-vector field) $\varphi =  \varphi_{\nu}dx^{\nu}$  satisfying $\partial_{\mu}\varphi_{\nu}-\partial_{\nu}\varphi_{\mu}=0$. In addition we assume  a real valued, positive {\em scalar field} $\phi$, in Riemann gauge (Jordan frame)  given by $\phi = C  e^{-\sigma}$ and scaling with   Weyl weight $-1$.  It plays a part in the gravitational structure of  our model.

 Being pure gauge, $\varphi$ can be  integrated away locally; then the Weylian metric
 acquires the form $(\tilde{g}, 0)$ and reduces to the (peudo-) Riemannian metric $\tilde{g}$. By obvious reasons this scale choice will be called the {\em Riemann gauge} of the Weylian metric. It corresponds to the Jordan frame of Jordan-Brans-Dicke (JBD) theory.\footnote{The usual terminology of  ``frame'' is ambiguous. Often it is used in the sense of choosing an orthonormal frame (tetrad); in JBD theory it indicates only a choice of scale, leaving coordinate choice and tetrad choice (if any) open. \label{fn gauge/frame}}
 Here the scalar field  can be written as 
\beq 
\tilde{\phi}= \phi_0 e^{-{\sigma}}  \, . \label{eq invariant potential}
\eeq
$\sigma$ is scale invariant by definition. Below we see that it plays the role of a potential for modifications of the kinematics of free fall in Einstein gauge (\ref{eq add acc 1}). We thus  call it the {\em invariant potential} of the scale connection.
 The scale connection does not entail a dynamical degree of freedom of its own; it arises from rescaling,  $\varphi_{\nu}= \partial_{\nu}\omega$  (see eq. (\ref{rescaled metrical data})). 
 In Einstein gauge/frame, in which the scalar field is scaled to a constant value, it expresses the dynamical content of the scalar field in the sense of  $\varphi_{\nu}= \partial_{\nu}{\sigma}$.

 The data in any other scale gauge arise from the Riemann gauge by  (length-) rescaling with a real valued function $\Omega=e^{\omega}$:
\beq g \doteq e^{2\omega}g , \qquad 
 \quad \varphi_{\nu} \doteq-\partial_{\nu} \omega \quad (\varphi \doteq - d \omega ),  \qquad
\phi \doteq \phi_0\, e^{-(\sigma + \omega)} \label{rescaled metrical data}
\eeq
Here $\doteq$ denotes equalities which are scale dependent. 
The Einstein gauge (also Einstein frame, but see fn. \ref{fn gauge/frame}) with metrical data  $(\hat{g},\hat{\varphi})$ and scalar field $\hat{\phi}$ is specified by the condition that the scalar field is scaled to a constant, $\hat{\phi}\doteq\phi_0$, thus:
\beq \hat{g} \underset{Eg}\doteq e^{-2 \sigma}g \, \qquad 
\hat{\varphi}_{\nu} \underset{Eg}\doteq \partial_{\nu}\sigma\, \qquad 
\hat{\phi} \underset{Eg}\doteq \phi_0 \quad \mbox{constant} \, , \label{eq Einstein gauge}
\eeq 
where $\underset{Eg}\doteq$ denotes equality in Einstein gauge; similarly $\underset{Rg}\doteq$ for  Riemann gauge. 

Partial derivatives are denoted as usual by $\partial$. For  covariant derivations we have to distinguish between the {\em Levi-Civita derivative} $_g{\hspace{-0.25em}}\nabla$ with regard to the Riemannian component $g$ of the Weylian metric, the {\em scale invariant} covariant derivative $\nabla$ with regard to the Weylian metric given by $(g, \varphi)$, and the {\em scale covariant derivative} $D$ of fields (which  themselves are  scale dependent). For  technical details see appendix \ref{Appendix IWG}.

It is convenient to introduce a sign symbol $\epsilon_X$ for scalar fields $X$, depending on the signature type of the scale covariant gradient $DX$:
\beq \epsilon_X = 
 \Big\{
{\mbox{\hspace{-2.5em}} +1 \quad \mbox{for}\quad DX \; \mbox{spacelike}  
\atop \; -1 \quad \mbox{for}\quad DX \; \mbox{timelike or null} }
\eeq 
Then the norm of a scale covariant gradient is  $|DX|=(\epsilon_X D_{\nu}XD^{\nu}X)^{\frac{1}{2}}$ and in particular  $|D\phi|\underset{Eg}\doteq \phi_0 |\nabla \sigma|$.

\subsection{\small Lagrange density \label{subsection Lagrangian}}
We assume a {\em scale invariant Lagrangian density} of the  form 
\beq \mathfrak{L}=L \sqrt{|\mathrm{det}\,g|}, \qquad L = L_H+L_{D\phi}+L_V  + L_{bar} \; ,
\eeq
with a gravitational term $L_H$, the kinetic and  potential terms of the scalar field $L_{D\phi}, L_{V(\phi)}$, and a {\em  matter term} $L_{bar}$ which we do not specify here. It serves  as a placeholder for the classical action of  baryonic matter. In order to cancel the scale weight of the volume element $ \sqrt{|\mathrm{det}\,g|}$, all contributions $L_{X}$  have to be given in scale covariant form   of Weyl weight $w(L_X)=-n=-4$. For $L_m$ one has to introduce  appropriate scaling conventions for its constituent fields without assuming a direct coupling to $\phi$. 

The {\em gravitational action} is similar to JBD-theory (in Riemann gauge the two are even equal):
\beq L_H= \frac{1}{2} (\xi \phi)^2 R \, , \qquad \mbox{with} \quad \xi=\frac{E_{P}}{\phi_0}\, , \label{eq L-H}
\eeq
where $R$ denotes   the {\em Weyl geometric scalar curvature}  (scale covariant of weight $w(R)=-2$, see app. \ref{Appendix IWG}, \ref{Appendix formulas}).  $\xi$ is a {\em hierarchy factor} between the Planck scale energy $E_P$ and the energy level of the scalar field, indicated by $\phi_0$.

The scale weight condition 
\beq L_{V} = - V \qquad \mbox{with} \quad w(V)=-4\; 
\eeq
constrains the form of $V$ to   a {\em quartic} monomial in the scalar field,
\[ V=V(\phi) = \frac{\lambda_4}{4}\phi^4 ,\]
 or to a {\em biquadratic} coupling of $\phi$ to the norm  $h$ of the Higgs field ($h^2=  \Phi^{\dag}\Phi$), with or without separate quartic term for $\phi$:\footnote{Similar to \citep{Shaposhnikov/Zenh"ausern:2009}.} 
\beq V(\phi,h) = \frac{\lambda}{4} \left( (h^2 - (\eta \phi)^2) \right)^2 \quad [+\frac{\lambda_4}{4}\phi^4 ] \; \label{eq biquadratic potential}
\eeq
$\eta = \frac{v}{\phi_0}$  is a new hierarchy factor between the  electroweak energy level $v \approx 246\, GeV$  and the energy of the scalar field $\phi$. With (\ref{eq biquadratic potential})  the  gravitational scalar field $\phi$ is able to enter the {\em Higgs portal}  in a moderate form (via the potential term only). In the case $\eta \sim 1$ one may expect $\lambda_4=0$.\footnote{In this case a  hierarchy factor $\xi'$   (different from $\xi$) between the MONDian constant $a_0$  and $\phi_0$ has to be introduced in the $L_{D\phi^3}$ eqs. (\ref{eq LDPhi3 gen}). \label{fn Higgs portal}}

The {\em kinetic term} $ L_{D\phi}$ of the model superimposes three terms:
\beq L_{D\phi} =  L_{D\phi^2} +  L_{D\phi^3} +  L_{D^2\phi} \label{eq L-Dphi} \, ,
\eeq 
where  $L_{D\phi^2}$, 
see (\ref{eq L-Dphi2}), denotes the usual quadratic kinetic term with coefficient $\alpha$ similar to JBD  written in scale covariant form,
$ L_{D\phi^3}$ (\ref{eq LDPhi3 gen}, \ref{eq LDphi3 Eg}) with coefficient $\beta$ is the cubic term adapted from the RAQUAL Lagrangian 
  \citep{Bekenstein/Milgrom:1984}, and $L_{D^2\phi}$ (\ref{eq L-bras}) with coefficient $\gamma$ the second order term  from  the Brazilian approach  to gravity mentioned above  \citep{Novello/Oliveira_ea:1993,Oliveira/Salim/Sautu:1997}.\footnote{The authors of \citep{Novello/Oliveira_ea:1993,Oliveira/Salim/Sautu:1997} assume a breaking of scale symmetry to the Einstein gauge, which is here avoided at the level of the general Lagrangian.} All terms are rewritten in scale covariant form (weight $-4$) with scale covariant derivatives $D$  in a general scale gauge and in Einstein gauge: 
  \beqarr   L_{D\phi^2} &=& - \frac{\alpha}{2} D_{\nu}(\xi \phi )  D^{\nu}(\xi \phi ) = - \epsilon_{\phi} \frac{\alpha}{2}|D(\xi \phi)|^2 \, \label{eq L-Dphi2} \\
 &\underset{Eg} \doteq & -\frac{\alpha}{2} (\xi \phi_0)^2 \, \partial_{\lambda}\sigma \partial^{\lambda}\sigma \nonumber \\
  L_{D\phi^3} &=& - \epsilon_{\phi}\frac{2}{3} \beta \phi^{-2} \ [ \epsilon_{\phi} D_{\nu}(\xi \phi )  D^{\nu}(\xi \phi )]^{\frac{3}{2}}  =-\epsilon_{\phi} \frac{2}{3} \beta \phi^{-2} |D(\xi \phi)|^3 \label{eq LDPhi3 gen} \\
   &\underset{Eg}\doteq& - \epsilon_{\sigma} \frac{2}{3} \beta (\xi \phi_0)^2 (\xi^{-1} \phi_0)^{-1} | \nabla \sigma|^3 \label{eq LDphi3 Eg} \\
   L_{D^2\phi} &=&  \frac{\gamma}{2} (\xi \phi) D_{\nu}D^{\nu}(\xi \phi) = \frac{\gamma}{2} \xi^2  \phi \, D_{\nu}D^{\nu} \phi \label{eq L-bras} \\
 &\underset{Eg}\doteq& - \frac{\gamma}{2} (\xi \phi_0)^2 (\nabla_{\nu}\partial^{\nu} \sigma - 3 \partial_{\nu} \partial^{\nu} \sigma ) \; \nonumber \\
 &\underset{Eg}\doteq& - \frac{\gamma}{2} (\xi \phi_0)^2 (_g{\hspace{-0.25em}}\nabla_{\nu}\partial^{\nu} \sigma +\partial_{\nu} \partial^{\nu} \sigma ) \; \nonumber
  \eeqarr

We basically assume two regimes in which the scalar field is governed by different Lagrangians, the   {\em eEG regime} (effectively Einstein gravity) in which the scalar field underlies a   JBD Lagrangian with a large coefficient $\alpha  \geq 10^5$ which leads effectively   to the dynamics of Einstein gravity, and the {\em MG} (modified gravity) {\em regime} with  a MOND-like dynamics. 
The Lagrangian of the MG regime  is ``switched on'' under specific conditions for the scalar curvature;  or the other way round, the MG regime is switched off if the gradient  of the scalar field surpasses a critical value roughly an order of magnitude below the MOND constant $a_0$. A physical explanation of this behaviour  of the present Larangian may result from a superfluid hypothesis  of an Einstein-Bose condensate 
 similar to   theory of  J. Khoury and L. Berezhiani \citep{Berezhiani/Khoury:2016,Berezhiani/Famaey/Khoury:2017,Berezhiani/Khoury:2019}. According to these authors the superfluid phase is suppressed for large phonon gradients of the superfluid which leads to a gradual transition to Newton/Einstein gravity with an increasingly smaller amount of superfluid phase of the condensate \citep[\S 5]{Berezhiani/Khoury:2015}.\footnote{Berezhiani/Khoury assume also a fading out of the superfluid phase for low pressures, i.e. far beyond galaxies, leading to  a phase with usual dark matter already at the cluster level. This is not the case for the present approach, but may be considered at the level of voids.}

\beq L_{D\phi} \;  =\; L_{D\phi}(\alpha, \beta, \gamma)\quad \mbox{with}\quad
 \Bigg\{
{\mbox{\hspace{-0em}} \alpha \gg1, \; \; \beta = \gamma = 0 
\quad \;  \mbox{in the eEG regime} 
\atop  \alpha=-6,\, \beta, \gamma \neq 0 \quad \hspace{1em} \mbox{in the MG regime}  } \label{eq LDphi}
\eeq 
In the sequel we find  a preference for setting $\beta=6$ (see (\ref{eq beta=6})) and reasons for assuming $\gamma=4$ (Result b) on p. \pageref{result b}). So we can put the  Lagrangians 
\[ L_{eEG} =L_{D{\phi}}(\alpha_1,0,0)\; , \qquad
 L_{MG}=L_{D_{\phi}}(-6,6,4)
 \]  (with some $\alpha_1\gg 1$) for the effective Einstein  and the modified gravity regimes respectively.
 
 At the moment there are no reasons for assuming a specific law for the {\em intermediate (im) region}  between the domains of eEG and MG. One may like, however, to postulate a transition function $\chi_{im}$ with  $\chi_{im}(x)=1$  in the MG region, $\chi_{im}(x)=0$  in the eEG region and a smooth transition, such that in the intermediate region
 \beq L_{im}= (1- \chi_{im})L_{eEG}+ \chi_{im}L_{MG}\, . \label{eq intermediate region}
 \eeq 
 This is at least a formal device for shunting out our  ignorance (even at the phenomenological level) of the physics in the intermediate region.

In the eEG regime we thus have $L_{D\phi}=L_{D\phi^2}$ with  $\alpha > 10^5$ which makes it compatible with Einstein gravity in the solar system  \citep{Will:LivingReviews};  in the MG regime all three contributions to $L_{D\phi}$ are switched on.   In the MG regime  a fractional power $\frac{3}{2}$ of the  quadratic term is turned on. According to Khoury/Berezhiani that is similar to what can be expected in the superfluid phase of a bosonic condensate \citep{Berezhiani/Khoury:2019}.  Fractional powers  (with exponent $\frac{3}{2}$)   of quadratic kinetic terms are typical for the  known Lagrangian field theories with MOND-like phenomenology (among them in particular the original RAQUAL  approach \citep{Bekenstein/Milgrom:1984}, TeVeS \citep{Bekenstein:2004},  the superfluid theory of Khoury/Berezhiani \citep{Berezhiani/Khoury:2016} and Hossenfelder's covariant version of ``emergent'' gravity \citep{Hossenfelder:2017}).  $L_{D^2\phi}$ contributes  considerably to the energy-momentum of the scalar field and leads to an important difference to the original RAQUAL approach. According to our hypothesis (\ref{eq LDphi})  is switched on in the MG regime  together with the cubic kinetic term, while the coefficient of the ordinary quadratic term is shifted to  conformal coupling, i.e. to a gravitationally inert state.

At the end of section \ref{subsection Newton-Milgrom} we come back to distinguishing criteria between  the different regimes in the context of a weak field  (Newton-Milgrom) approximation. The criteria given there are based on rough estimates on the validity region for a MOND-like dynamics. A theoretical underpinning for such a separation of regions is still  conjectural, but there are arguments for the appearance of  ``kinematic screening'' of scalar fields in the theory of superfluids.\footnote{Berezhiani/Khoury argue in  \citep[p. 4]{Berezhiani/Famaey/Khoury:2017}: ``The EFE (external field effect, ES) is an example of a more general phenomenon in scalar field theories known as kinetic screening [\ldots].
In theories with gradient interactions, non-linearities in the scalar field gradient -- the scalar acceleration -- can
result in the suppression of the scalar field effects and the
local recovery of standard gravity. See [75] for a review.''  ([75] refers to our \citep{Jouyce/Khoury_ea:2015}.)} 
A  better consolidated   theory would probably formulate the distinguishing criteria for the two regimes in terms of the  scalar field gradient.

\subsection{\small Einstein equation and its energy momentum terms \label{subsection Einstein equation}}
In order to take full advantage of the scaling symmetry of the Lagrangian, the variational derivatives are dealt with in a scale covariant framework  like in  other gauge symmetric theories \citep[p. 524ff.]{Frankel:Geometry}. The  scale covariant Euler-Lagrange  equation with regard to $g$, 
\beq \frac{\partial \mathfrak{L}}{\partial  g^{\mu\nu}} - D_{\lambda} \frac{\partial  \mathfrak{L}}{\partial(D_\lambda  g^{\mu\nu})} \, \label{eq variation g}
\eeq 
can be calculated in any scale gauge if in the end the result  is rewritten in  scale covariant form. In our case,  the integrable Weyl geometric  context makes the  calculation easy. In most cases, one can go to  Riemann gauge and use the fact that all derivative operators mentioned at the end of sec. \ref{subsection Framework} are equal,  $D\underset{Rg}\doteq \nabla \underset{Rg}\doteq \, _g\hspace{-0.25em}\nabla$.  

After multiplying the result with $(\xi \phi)^{-2}$ we arrive at the following  {\em scale invariant}  {\em Einstein equation}: 
\beq G= Ric - \frac{R}{2}g = (\xi \phi)^{-2} T^{(bar)}+ \Theta^{(H)} + \sum_X \Theta^{(X)} \label{eq Einstein equation}
\eeq
Here $Ric=(R_{\mu\nu})$,  $R$ and $G=(G_{\mu\nu})$ denote the {\em Weyl geometric} Ricci tensor, scalar curvature, and Einstein tensor respectively   (see appendix \ref{Appendix IWG}).  $Ric$ is scale invariant by definition, $w(R)=-2$ , $w(g)=2$. Therefore also the Weyl geometric Einstein tensor is  scale invariant. Similar weight counts hold for all terms on the right hand side (r.h.s.) of eq (\ref{eq Einstein equation}).
This equation  also holds without assuming integrability of the Weylian scale connection  \citep{Smolin:1979,Drechsler/Tann,Blagojevic:Gravitation}).

The Weyl geometric Einstein tensor in scale gauge $(g,\varphi)$ is   the sum of a term $_gG$ due to the Riemannian part of the  metric $g$ and well  known from Einstein gravity, and an expression $_{\varphi}\hspace{-0.1em}G$ containing the contributions of the scale connection (appendix \ref{Appendix IWG}), 
\beq 
G=\, _gG+ _{\varphi}\hspace{-0.1em}G \label{eq decomposition of G}
\eeq

The energy momentum tensor of classical matter 
\[ T^{(bar)}_{\mu \nu}= - \frac{2}{\sqrt{|g|}}\frac{\delta \mathfrak{L}_{bar}}{\delta g^{\mu\nu}} \, \]
calculated according to (\ref{eq variation g}) 
is scale covariant with weight $-2$ which cancels  against $w(\phi^{-2})$.  Similarly  the 
 $\Theta^{(X)}$ denote the scale invariant contributions to the energy momentum tensor of the scalar field, up to coefficient:
 \[ \Theta^{(X)}= - (\xi \phi)^{-2}\frac{2}{\sqrt{|g|}}\frac{\delta \mathfrak{L}_{X}}{\delta g^{\mu\nu}} = - (\xi\phi)^{-2}\left(2 \frac{\partial L_X}{\partial g^{\mu\nu}}-L_X\, g_{\mu\nu}\right)
 \]  
Here $X$ is used as a dummy index for the  constituents of the scalar field with the appropriate summation domains, i.e. $X\in \{  D\phi^2,V \}$ in the eEG region, and $X\in \{ D\phi^2 , D\phi^3 , D^2\phi  V \}$ in the MG regime  with $\alpha=-6$.

 $\Theta^{(H)}$ is the variational contribution of $\phi$   due to the non-minimal coupling in $\mathfrak{L}_H$. It is well known also in JBD theory \citep{Fujii/Maeda,Capozziello/Faraoni}. Written in terms of the scale covariant differentiation operator $D$ of Weyl geometry  it is \citep{Drechsler/Tann,Tann:Diss}: 
 \beqarr \Theta_{\mu\nu}^{(H)} &=& \phi^{-2}\left(D_{(\mu}D_{\nu)}\phi^2 -D_{\lambda}D^{\lambda}\phi^2\, g_{\mu\nu} \right) \label{eq Theta (H)} \\
  &=& 2 \phi^{-2} \left( D_{\mu}\phi D_{\nu}\phi + \phi D_{(\mu}D_{\nu)}\phi - (\phi D_{\lambda}D^{\lambda}\phi + D_{\lambda}D^{\lambda}\phi)\, g_{\mu\nu} \right) \nonumber  \\
  \mbox{Using} && \mbox{\hspace{-2.5em} the calculation in app. \ref{Appendix formulas} this is in Einstein gauge} \nonumber \\
  &\underset{Eg} \doteq& -2 \, _g{\hspace{-0.25em}}{\nabla}_{\hspace{-0.2em}(\mu} \partial_{\nu)} \sigma + 8\, \partial_{\mu}\sigma   \partial_{\nu}\sigma - 2\, (_g \square \, \sigma + \partial_{\lambda}\sigma  \partial^{\lambda}\sigma  ) \,  \nonumber
 \eeqarr
 Here 
  \beq _g\square  = - _g\hspace{-0.25em}\nabla_{\hspace{-0.2em}\lambda} \partial^{\lambda} \label{eq d'Alembert op}
 \eeq 
denotes the d'Alembert operator with regard to the Levi-Civita connection $_g\hspace{-0.25em}\nabla$.

The interpretation of $\Theta^{(H)}$ in the literature varies; some authors consider it as a  geometrico-gravitational contribution to the Einstein equation and put it on its  left hand side (l.h.s.), others see it as part of the energy-momentum of the scalar field.\footnote{This was done in \citep{Scholz:MONDlike}.} In section \ref{subsection MOND} we come  back to it  in  our context.

The other  energy expressions (most of them in scale invariant form) are 
\beqarr
\Theta^{(D\phi^2 )}_{\mu\nu} &=& (\xi \phi)^{-2} (\alpha \xi^2 D_{\mu}\phi D_{ \nu}\phi+ L_{D\phi^2}\, g _{\mu \nu}  ) \label{eq Theta Dphi2} \\
 &= & \alpha ( \partial_{\mu} \sigma \partial_{\nu} \sigma - \frac{1}{2}\partial_{\lambda} \partial^{\lambda} \sigma  \, g_{\mu\nu} ) \nonumber \\
 \Theta^{(D\phi^3 )}_{\mu\nu} &=& 2 \beta \xi \phi^{-4}\, |D\phi| D_{\mu}\phi D_{\nu}\phi + (\xi \phi)^{-2 }L_{D\phi^3}\, g_{\mu \nu} \qquad \\ 
  & \underset{Eg}\doteq & 2 \beta (\xi^{-1} \phi_0)^{-1}\left( |\nabla \sigma |\partial_{\mu} \sigma \partial_{\nu} \sigma - \frac{\epsilon_{\sigma}}{3} |\nabla \sigma|^3\, g_{\mu \nu} \right) \nonumber \\
   \Theta^{(D^2\phi)}_{\mu\nu} &=&  \gamma \phi^{-1} \left(- D_{\mu}D_{\nu} \phi  + \frac{1}{2}D_{\lambda}D^{\lambda}\phi \, g_{\mu\nu} \right) \label{eq Theta bras} \\
    & \underset{Eg}\doteq &  -\gamma \left( _g\hspace{-0.25em} \nabla_{\hspace{-0.15em}\mu}\partial_{\nu} \sigma  -3\,  \partial_{\mu} \sigma \partial_{\nu} \sigma \right)  + \frac{\gamma}{2} \left(\, _g\hspace{-0.2em}\nabla_{\hspace{-0.25em}\lambda} \sigma^{\lambda} -3\, \partial_{\lambda} \sigma \, \partial^{\lambda} \sigma  \right) \, g_{\mu\nu} \nonumber \\ 
   \Theta^{(V)}_{\mu\nu} &=&    (\xi \phi)^{-2}L_V\, g_{\mu\nu}  
\eeqarr
Tracing the Einstein equation and multiplying it with  $ -(\xi \phi)^2$ leads to 
\beq  - 2 L_H - tr\, T^{(bar)} + (\gamma +6)\xi^2 \phi D_{\lambda}D^{\lambda}\phi  + (\alpha +6)\xi^2 D_{\lambda}\phi D^{\lambda}\phi- L_{D\phi^3} - 4 L_V = 0 \, . \label{trace Einstein eq}
\eeq 
Here all possible contributions of the scalar field are included. In the JBD domain one has to set  $\gamma = 0$ and $\beta =0$, in the MG regime $\alpha=-6$.

\subsection{\small Scalar field equation \label{subsection scalar field equations}}

 The scale covariant variation with regard to $\phi$, $  \frac{\delta L}{\delta  \phi} =
\frac{\partial L}{\partial  \phi} - D_{\lambda} \frac{\partial  L}{\partial(D_{\lambda}  \phi)} $,
contains the partial contributions (see app. \ref{Appendix formulas} (\ref{eq variation LDphi3})): 
\beqa      \\
\frac{\delta L_{D\phi^2}}{\delta  \phi} &=& \alpha \xi^2 D_{\lambda}D^{\lambda}\phi\, \qquad \\ 
\frac{\delta L_{D\phi^3}}{\delta  \phi} &=& 2 \beta \,\xi^3 \phi^{-2}\, D_{\lambda}\left(|D \phi | D^{\lambda}\phi\right) + 4 \phi^{-1}\, L_{D\phi^3} 
\eeqa
In the second line we encounter  a scale covariant form of the  non-linear modification of the d'Alembert operator typical for relativistic MOND theories.  

For  $ L_{D^2\phi}$ it is recommendable to consider the Einstein gauge   (\ref{eq L-bras}). 
Because of 
\[ _g\hspace{-0.25em}\nabla_{\hspace{-0.2em}\lambda}  \partial^{\lambda}\, \sigma = \frac{1}{\sqrt{|g|}} \partial_{\lambda} (\sqrt{|g|} \partial^{\lambda} \sigma)
\] 
the second order derivative term of $ \mathfrak{L}_{D^2\phi}$ in Einstein gauge is a divergence expression
\[- \frac{\gamma}{2} (\xi \phi_0)^2 \partial_{\lambda} (\sqrt{|g|} \partial^{\lambda} \sigma) \, .
\] 
For the variation of $\phi$ (with fixed $g$) its integral 
 can be shifted to a boundary term outside the support of  $\delta \phi$  and  does not contribute to the Euler-Lagrange equation of the scalar field.\footnote{This has been noted  by the authors of \citep{Novello/Oliveira_ea:1993}.}
The same does not hold while varying   $g$.   
 For the variation $\delta \phi$  in Einstein gauge only the  term $-\frac{\gamma}{2} (\xi \phi_0) \partial_{\lambda} \sigma \partial^{\lambda} \sigma $ of (\ref{eq L-bras}) remains and leads to a second degree dynamcial equation for $\phi$, respectively $\sigma$.  Its scale covariant form  is 
 \[ L_{D^2\phi  reduced}= - \frac{\gamma}{2} (\xi \phi_0)^2 D_{\lambda} \sigma D^{\lambda} \sigma\, ;
 \]
it is of the same form as $L_{D\phi^2}$  (\ref{eq L-Dphi2}). In terms of $\phi$ we find
 \beq
\frac{\delta L_{D^2\phi}}{\delta  \phi} = \gamma \xi^2 D_{\lambda}D^{\lambda}\phi \, .
\eeq 
$L_H$ and $L_V$ are monomials in $\phi$ with $\frac{\delta \phi^k}{\delta \phi} = \frac{\partial \phi^k}{\partial \phi}= k \phi^{k-1}$. \\[0.3em]

After summing up and multiplying with $\phi$ we arrive at the {\em gross  scalar field equation} covering both regimes of (\ref{eq LDphi}):
\beq 2 L_H + (\alpha - \gamma)\,  \xi^2 \phi\, D_{\lambda} D^{\lambda}\phi + 4 L_{D\phi^3} + 2 \beta \xi^3 \phi^{-1}\, D_{\lambda}\left(|D \phi | D^{\lambda}\phi\right) + 4 L_V = 0
\eeq
Addition of the traced Einstein equation  (\ref{trace Einstein eq}) leads to the scale covariant, (net) {\em  scalar field equation}. In arbitrary scale gauge it is:
\beq 2 \beta \xi^3 \phi^{-1}\, D_{\lambda}\left(|D \phi | D^{\lambda}\phi\right) + (\alpha + 6)\xi^2  D_{\lambda}\phi D^{\lambda}\phi + (\alpha +6)\,  \xi^2 \phi\, D_{\lambda} D^{\lambda}\phi + 3 L_{D\phi^3} - tr\, T^{(bar)}=0 
\eeq 
In the {\em MG regime} ($\alpha = -6$) this boils down to
\[  2 \beta \xi^3 \phi^{-1}\, D_{\lambda}\left(|D \phi | D^{\lambda}\phi\right) + 3 L_{D\phi^3} =  tr\, T^{(bar)} \, . 
\] 
 Taking account of  $D_{\lambda}(|D\phi |D^{\lambda}\phi) \underset{Eg} \doteq  -\phi_0^2 D_{\lambda}(|\nabla \sigma|\partial^{\lambda}\sigma)$,  $\phi \underset{Eg} \doteq \phi_0$, and  for a perfect fluid energy tensor with energy density $\rho_{bar}$ and pressure $p_{bar}$ this becomes  in Einstein gauge:
\[  D_{\lambda} \left(|\nabla \sigma | \partial^{\lambda} \sigma  \right) - \epsilon_{\sigma} |\nabla \sigma |^3 \underset{Eg} \doteq   \frac{1}{2}(\xi\phi_0)^{-2} \left((\beta \xi)^{-1}\phi_0 \right)\, (\rho_{bar} - 3 p_{bar}) \, 
\]
With eq.  (\ref{cubic cancelling})  of the  appendix,
\[   D_{\lambda} \left(|\nabla \sigma | \partial^{\lambda} \sigma  \right)   \underset{Eg} \doteq \, 
 _g\hspace{-0.2em} \nabla_{\hspace{-0.15em}\lambda} \left(|\nabla \sigma | \partial^{\lambda} \sigma  \right)+ |\nabla \sigma| \partial_{\lambda} \sigma  \partial^{\lambda} \sigma \,,
\] 
 the cubic terms  $|\nabla \sigma|^3$  cancel, and  we arrive at the scalar field equation in the MG regime:
\beq
 _g\hspace{-0.2em} \nabla_{\hspace{-0.15em}\lambda} \left(|\nabla \sigma | \partial^{\lambda} \sigma  \right) \underset{Eg} \doteq  \quad   \frac{1}{2}(\xi\phi_0)^{-2} \left((\beta \xi)^{-1}\phi_0 \right)\,\; (\rho - 3p)_{bar} \,. \label{eq covariant Milgrom equation}
\eeq 
Let us define the 
 {\em covariant Milgrom operator} $  _g\Delta_M$ in Einstein gauge, for  any scalar field $X$   as
\beq   _g\Delta_M X  :\underset{Eg} \doteq   \,  _g\hspace{-0.2em} \nabla_{\hspace{-0.15em}\lambda} \left(|\nabla X |\, \partial^{\lambda} X  \right) \, . \label{eq covariant Milgrom op}
\eeq 
For the flat metric and static fields $X$ this is the non-linear Laplace operator of the classical MOND theory $\Delta_M X = \nabla_j (|\nabla X|\, \partial^j X)$ (with the Euclidean $\nabla$ operator). We therefore call (\ref{eq covariant Milgrom equation}) the {\em covariant  Milgrom equation}.

In the {\em eEG regime} ($\beta=\gamma=0$), on the other hand,   we get
\[ (\alpha + 6)\xi^2 ( D_{\lambda}\phi D^{\lambda}\phi + \phi\,  D_{\lambda} D^{\lambda}\phi ) =  tr\, T^{(bar)} \, ,
\]
or
\[ D_{\lambda}D^{\lambda}\phi^2 = \frac{2}{\alpha+6}\xi^{-2}\, tr\, T^{(bar)} \, .
\]
In Einstein gauge and for fluid matter this is
\beq _g\square\, \sigma \; \underset{Eg} \doteq  \; \frac{2 (\xi \phi_0)^{-2}}{\alpha +6}\, (\rho-3p) \, ,
\eeq
the scalar field equation of JBD theory in the Einstein frame (cf. \citep[pp. 42, 72]{Fujii/Maeda}).
(Remember that $ _g\square$ denotes the covariant d'Alembert operator (\ref{eq d'Alembert op}).)

\subsection{\small Free fall trajectories \label{subsection trajectories}}
If we model the trajectories of test bodies by energy-momentum concentrated in arbitray small neighbourhoods of a timelike curve, like  in the Geroch-Jang approach to the geodesic theorem for Einstein gravity \citep{Geroch/Jang:1975}, it can be  shown  that   in integrable Weyl geometry (IWG)  test bodies  move  along timelike geodesics like in Einstein gravity. One basically   passes to the Riemann gauge and applies the ``classical'' Geroch-Jang theorem (appendix \ref{Appendix Geroch-Jang}). 

The scale invariant geodesics $\tilde{\gamma} (\tau)$ of IWG, 
\[ \frac{d}{d\tau}\dot{\tilde{\gamma}}^{\mu} + \Gamma^{\mu}_{\nu \lambda}\dot{\tilde{\gamma}}^{\nu}\dot{\tilde{\gamma}}^{\lambda} = 0
\]
can be expressed as the Levi-Civita geodesics of Riemann gauge ($\Gamma \underset{Rg} \doteq \, _g\Gamma$).  It is useful to introduce  also a scale covariant parametrization $\gamma(\tau)$ for the (timelike) geodesics such that $g(\dot{\gamma}, \dot{\gamma}) = -1$ in all scale gauges (a kind of scale dependent proper time parametrization). This means that one works with {\em scale covariant geodesics} for which the weight of the tangent vector field is  $w(\dot{\gamma})=-1$.
 
  In local coordinates with $x_0=t$ and spacelike indices   $i,j,k = 1, \ldots 3$ the geodesic equation for a scale covariant timelike geodesic has the same form as in Einstein gravity (cf. e.g. \citep[eq. 9.1.2]{Weinberg:Cosmology_1972}), but in our case the connection coefficients $\Gamma$ are  the  Weyl geometric ones  \citep[p. 4]{Scholz:Clusters}: 

\beqarr \label{equ of motion}
\frac{d^2 x^{j }}{dt^2} &= & 
- \Gamma^{j}_{00} +\Gamma^{0}_{00} \frac{dx^{j}}{dt}
 - 2 \Gamma^{j}_{0 i} \frac{dx^{i}}{dt}
-  \Gamma^{j}_{ i k} \frac{dx^{i}}{dt}\frac{dx^{k}}{dt}  \\
&  & + 2  \Gamma^{0}_{0 i } \frac{dx^{j}}{dt}\frac{dx^{i}}{dt}
+ \Gamma^{0}_{ i k } \frac{dx^{j}}{dt}\frac{dx^{i}}{dt}\frac{dx^{k}}{dt}\, 
\nonumber 
\eeqarr
 In the low velocity, weak field regime the equation of motion reduces to a form well known from Einstein gravity, $\frac{d^2 x^{j }}{dt^2} =  
- \Gamma^{j}_{00}$.  The  $\Gamma^{j}_{00}$ ($j=1,2,3$) are now the coefficients of the  Weyl geometric affine connection which differs from the Levi-Civita connection  of the Riemannian component $g$ by (\ref{eq varphi Levi-Civita}). In any scale gauge $(g,\varphi)$  different from Riemann gauge the coordinate acceleration $a$ of freely falling bodies takes up terms from the scale connection, in addition to the Levi-Civita contributions of the Riemannian part $g$ of the metric; in the low velocity, weak field case:
\beq a^{j}= \frac{d^2x^{j}}{d\tau^2} \approx - \Gamma^{j}_{00} =  - _g\hspace{-0.1em}\Gamma^{j}_{00} - _\varphi\hspace{-0.15em}\Gamma^{j}_{ 00}  =  a^j_g +  a^j_{\varphi} \label{acc 1}
\eeq

To the well known metrical acceleration $a^j_g = - _g\hspace{-0.1em}\Gamma^{j}_{oo}  $ known from Riemannian geometry  a  component $ a^j_{\varphi}=- _\varphi\hspace{-0.05em}\Gamma^{j}_{ 00} = g_{00}\,\varphi^j$ induced from the Weylian scale connection is  added (cf. eq. (\ref{eq varphi Levi-Civita}). 
We call these terms the {\em Riemannian acceleration} and the {\em additional acceleration} due to the scale connection, respectively the scalar field. From the Riemannian (not the Weyl geometric) point of  view the latter appears as a ``{\em non-metrical}'' contribution to the acceleration. 
In Einstein gauge the additional scalar field acceleration for our Lagrangian  is (cf.  
 eqs. (\ref{eq Einstein gauge}), (\ref{eq varphi Levi-Civita}))
\beq a^j_{\varphi} =g_{00}\,\varphi^j \underset{Eg} \doteq   g_{00}\, \partial^j \sigma \, . \label{eq add acc 1}
\eeq 
For the dynamics of freely falling test bodies the exponent $\sigma$ of the scalar field in Riemann gauge functions  like an additional gravitational potential. In this respect our approach  clearly is a {\em modified gravity model}.

\subsection{\small Einstein gauge and the measured values of observable quantities \label{subsection Einstein gauge}}
Like in  JBD theory one may wonder  in IWG gravity which scale gauge  expresses  the measured values of observable quantities most directly. This may be understood as the question for a {\em bridge rule}  allowing to  connect the theoretical  (scaling) quantities with the (non-scaling) measured  values  of empirical quantities. In principle it is possible to formulate such a bridge rule without breaking    the scale symmetry just by  introducing  the scale invariant observable quantity $\hat{X}= \phi^{w(X)}X$  for any scale covariant field quantity $X$ of weight $w(X)$. The scale invariant observable of the scalar field is then  $\hat{\phi}=1$. This boils down, up to a constant, to considering the Einstein gauge.
 This, and the criterion of a best link to Einstein gravity, leads to the\\[0.5em]
\noindent
{\bf Bridge rule}: The theoretical values of field quantities of IWG gravity are to be identified  with the corresponding  empirical  values (basically astronomical and astrophysical ones) by going to the {\em Einstein gauge}.\\[-0.2em]

In this gauge the coefficient of the Hilbert term (\ref{eq L-H}) is a constant which can be identified with  Einstein's gravitational constant, 
\beq (\xi \phi)^2 \underset{Eg}\doteq (\xi \phi_0)^2  = (8 \pi G)^{-1} [\hbar c^5] \, . \label{eq 8 pi G}
\eeq
This leads to the closest possible  agreement with Einstein gravity in the eEG regime and to Einstein gravity as an exact limit for $\alpha \rightarrow \infty$.  Moreover, assuming the biquadratic potential coupling to the Higgs field (\ref{eq biquadratic potential}), the expectation value $h_0^2 =\Phi_0^{\dag}\Phi_0$ of the  Higgs field in the  potential minimum (the rest state $\Phi_0$ of the Higgs field)  also becomes spacetime independent in the Einstein gauge and acquires its experimental value $h_0^2 = v^2$ ($v\approx 246\, GeV$). One may  like to turn this argument round and argue that the biquadratic coupling to the Higgs field {\em breaks  the scale symmetry} of the theory. But this leads to a different discussion. 

Moreover,  we see in sec. \ref{subsection MOND}  that the choice of parameters such that
\beq (\beta \xi)^{-1} \phi_0  \underset{Eg}\doteq  a_0 \, [c^{-1}\hbar] \, \label{eq a_0}
\eeq
with the MOND constant $a_o\approx \frac{1}{6}H_0 [c]$ leads to an agreement with MOND kinematics in the deep MOND regime (but has interesting differences  with regard to gravitational lensing). 
Then the covariant Milgrom equation (\ref{eq covariant Milgrom equation}) becomes
\beq   _g\Delta_M \sigma  \underset{Eg} \doteq   \, 4\pi G\,a_0\, (\rho-3p)^{(bar)}
\eeq 
Choosing $\beta=6$ fixes the value of $\phi_0$ such that
\beq \xi^{-1}\phi \underset{Eg}\doteq \xi^{-1}\phi_0 = H_0 \label{eq beta=6}
\eeq 
Then $\phi_0$  is  the geometric mean between the smallest and the largest physically meaningful energy levels we know 
$ \phi_0\, [\hbar] = \sqrt{H_0 [\hbar] \, E_{P} }
$. 
 For exploring a possible connection to the Higgs portal one may prefer to  relax this constraint for $\phi_0$ by a different choice of coefficients in    (\ref{eq LDphi3 Eg}).

\section{\small MG regime and Newton-Milgrom approximation  \label{section MG}}

\subsection{\small The modification of Einstein gravity in the MG regime \label{subsection MG}  }
The equations (\ref{eq Theta (H)}, \ref{eq Theta Dphi2}) and (\ref{eq varphi G})  show that in Einstein gauge and for conformal coupling  of $L_{D\phi^2}$ ($\alpha=-6$), the scale connection part of  the Einstein tensor    compensates two terms of the energy momentum expression on the r.h.s. of (\ref{eq Einstein equation}):\footnote{This cancelling has been overlooked in \citep{Scholz:MONDlike,Scholz:Clusters}. The tensor $\Theta^{(H)}$ contains second order derivative terms comparable to $\Theta^{(D^2\phi)}$ of the present approach (see below, eq. (\ref{eq Theta eff})). The  dynamics of \citep{Scholz:MONDlike,Scholz:Clusters} seemed essentially the same as here (corresponding to $\gamma = 2$) -- but  at the price of a flawed derivation. See appendix \ref{Appendix Flaws} \label{fn overlooked}
 }
\beq  _{\varphi}\hspace{-0.1em}G \underset{Eg}\doteq \Theta^{(H)} + \Theta^{(D\phi^2 )} \,  \label{eq conformal Theta compensation}
\eeq 
In Einstein gauge and for the MG regime  eq. (\ref{eq Einstein equation})  reduces to an equation for the 
Riemannian component $g$ of the Weylian metric $(g,\varphi)$: 
 \beq
  _g\hspace{-0.1em}G \underset{Eg}\doteq  8 \pi G\, T^{(bar)} + \Theta^{(RAQ)} \; [+  \Theta^{(V)}] \,  \label{eq Einstein eq MG}
 \eeq
This is  a classical Einstein equation for $g$  with the r.h.s. enhanced by an energy momentum term $T^{(\phi)}$ of the scalar field,
\beqarr  T^{(\phi)} &\underset{Eg}\doteq& (8 \pi G)^{-1}\, \Theta  \nonumber \\
\mbox{with }\qquad \qquad  \Theta   &\underset{Eg}\doteq& \Theta^{(D^2\phi)} + \Theta^{(D\phi^3)}\; [+  \Theta^{(V)}]  \; .  \label{eq Theta Eg}
\eeqarr

The covariant Milgrom equation (\ref{eq covariant Milgrom equation}) and (\ref{eq 8 pi G}, \ref{eq a_0}) teach us that the  scalar field has only  baryonic matter as its source:
\[  _g\Delta_M\, \sigma \quad  \underset{Eg} \doteq  \quad (4\pi G)\, a_0\, (\rho - 3p)_{bar}
\]
This is an intriguing observation: The scalar field dynamics is sourced by baryonic matter only, while  the Riemannian component of the metric has  baryonic matter and the scalar field energy-momentum for its source. In this sense one may  consider the scalar field   as having also the  character of  {\em dark matter}  in addition to its being a part of the gravitational structure (section \ref{subsection trajectories}), although with a peculiar energy-momentum tensor $T^{(\phi)}$ (see sec. \ref{subsection enhanced momentum}).

We finally get a twofold modification of Einstein dynamics. The Riemannian acceleration $a_g$ is  not only due to the baryonic matter,  it contains a contribution from the scalar field, which may vanish for special field constellations,  see below eq. (\ref{eq trace T phi}). Moreover, the total acceleration (\ref{acc 1}) has an additional  component $a_{\varphi}$ due to the scalar field. Both together determine free fall with the total acceleration $a= a_g + a_{\varphi}$. The acceleration (\ref{acc 1}) refers to a  weak field, low velocity approximation, but  in the general case a similar modification of Einstein gravity holds.

\subsection{\small  Newton-Milgrom  approximation \label{subsection Newton-Milgrom}}
The Newton approximation of (\ref{eq Einstein eq MG}) uses a  weak field approximation for the  Riemannian component of the metric  $g$ in Einstein gauge  of the form 
\beq  g \underset{Eg}\doteq \eta + diag\, (h_{00}, h_{11}, h_{22}, h_{33}) \qquad \mbox{with} \quad \eta = diag\, (-1,1,1,1) \label{eq eta + h} \eeq
in which the $h_{jj} \approx 0$  can be neglected, while   $h_{00}=-2 \Phi_N$ plays the role of the Newton potential \citep[p. 153f.]{Carroll:Spacetime}. The approximation to the spatial part of the metric is  Euclidean. The first order approximation of $\Gamma$ leads to  $_g\Gamma_{00}^j \approx - \frac{1}{2} \eta^{jj}\partial_j h_{00}$. This motivates to set 
\beq \Phi_N = - \frac{1}{2}h_{00} \, ;
\eeq
then the  Riemannian acceleration of (\ref{acc 1}) acquires the form 
 \beq   a_g = - \nabla \Phi_N \, 
 \eeq 
with  the Euclidean gradient operator  $\nabla$, like in Newton gravity. 

For the  Riemannian component of the Ricci tensor we get, at first order in $h$  \citep[p. 158]{Carroll:Spacetime}, 
\beq _gR_{00} \approx- \nabla^2 h_{00} \, .
\eeq 
The Einstein equation (\ref{eq Einstein eq MG}) solved for the Ricci term is 
\beq _gRic = 8\pi G \, \left(T - \frac{1}{2} tr\, T \, g \right) \qquad \mbox{with} \quad T = T^{(bar)}+T^{(\phi)}\, , \label{eq Ricci term solved Einstein equ}
\eeq 
($tr$ the trace operator). Because of  $\; - tr\, T = \rho-\sum_1^3 p_j\;$ its $00$-component  leads to the Poisson equation for the Newton potential:
\beq  \nabla^2 \Phi_N \approx  4\pi G \left( \rho^{(bar)}+ \sum_1^3 p_j^{(bar)}  + \rho^{(\phi)} + \sum_1^3 p_j^{(\phi)} \right) \; \qquad \label{eq Poisson equation}
\eeq 
This completes the alignment with Newton dynamics. Here 
$\rho^{(X)}$ and $p_j^{(X)}$ denote the density and pressure components of baryonic matter and the scalar field energy-momentum respectively,  where  $X \in \{ bar, \phi\} $.
 In the Newton approximation the contribution of the scalar field to the r.h.s. of the Einstein equation is expressed  by  density and pressure terms analogous to those of  classical matter. But here we can no longer expect pressure components $p_j=p$ independent of the coordinate direction like for a classical fluid.
  
The Newton potential and Newton acceleration of the baryonic matter source alone are:
\beq  \nabla^2 \Phi_N^{(bar)} = 4\pi G \left( \rho^{(bar)} + \sum_1^3 p_j^{(bar)}   \right) \;   \qquad \quad a_N^{(bar)}=-\nabla \Phi_N^{(bar)} \label{eq baryonic Poisson equation}
\eeq
Analogously  for the scalar field contribution:
\beq \hspace*{-7em} \nabla^2 \Phi_N^{(\phi)} = 4\pi G \;(2 \rho^{(\phi)} + tr\, T^{(\phi)}) \; ,  \qquad \quad a_N^{(\phi)}=-\nabla \Phi_N^{(\phi)} \label{eq phi Poisson equation}
\eeq
Here 
\beq \rho_N^{(\phi\; \mbox{st})} = 2 \rho^{(\phi)} + tr\, T^{(\phi)}=\rho^{(\phi)}+\sum_1^3 p_j^{(\phi)} \label{eq mass density equivalent}
\eeq 
is the   {\em Newton-Poisson source term} (st) of the scalar field.

In a weak field regime of a quasi-static scalar field (i.e. one  one with relativistically  slow velocities/time dependence) considered in  the corresponding  Newton approximation,  the covariant Milgrom equation (\ref{eq covariant Milgrom equation}) gets a form  like in the  classical (i.e. Euclidean/Newtonian) case of the deep MOND dynamics
 \citep{Milgrom:1983} 
\beq \nabla \cdot \left(|\nabla \sigma | \nabla \sigma \right) = (4\pi G)\, a_0\; (\rho^{(bar)} - 3p^{(bar)}) \,. \label{eq Milgrom equation}
\eeq
Here $\nabla$ denotes the Euclidean gradient operator   and ``$\cdot$'' the Euclidean scalar product. The combination of (\ref{eq Poisson equation}) and (\ref{eq Milgrom equation}) will be called a {\em  Newton-Milgrom approximation}  of the relativistic dynamics given by  (\ref{eq Einstein equation})/(\ref{eq Einstein eq MG}) and (\ref{eq covariant Milgrom equation}). 

From the empirical evidence acquired in the framework of the classical MOND theory we may conclude that the onset of the MG regime, expressed in quantities of the Newton approximation, occurs in a region in which the norm of the baryonic Newton acceleration comes close to the order of magnitude of the  MOND constant $a_0 [c]$, in short  $|\,a^{(bar)}| \leq 10^{2k} a_0$, while the JBD (here Einstein/Newton) regime sets in roughly for $|\,a^{(bar)}| \geq   10^{2(k+l)}$ with, say, $k=l=1$.\footnote{We often omit factors $c$ and $\hbar$ in theoretical calculations and plug them in as soon as we approach empirical data.} 
These conditions  serve as  provisional {\em distinguishing criteria between the different regimes} of the scalar field Lagrangian (see end of sec. \ref{subsection Lagrangian} and sec. \ref{subsection MOND}).

 In principle the scalar field contributes two terms $a_N^{(\phi)},\, a_{\varphi}$  according to eqs. (\ref{eq phi Poisson equation}) and  (\ref{eq add acc 1}) 
to the total acceleration in the MG regime: 
\beq a_{tot} = a_R + a_{\varphi} =  a_N^{(bar)} + a_N^{(\phi)} + a_{\varphi}\,. \label{eq a-tot}
\eeq 
But we see in a moment (\ref{eq trace T phi}) that for simple systems the middle term $ a_N^{(\phi)}$ vanishes.  
From (\ref{eq add acc 1}) we know that the  scale connection term  is here simply 
\beq a_{\varphi} =-\nabla \sigma \, . \label{eq a phi Milgrom approximation}
\eeq
In order to compare it with the Newtonian one of the scalar field $ a_N^{(\phi)}$ we have a closer look at the density and pressure terms of the scalar field in a given weak field regime.

\subsection{\small The    scalar field energy-momentum in the Milgrom approximation \label{subsection enhanced momentum}}

To get an impression of the order of magnitude relations of different entries of $\Theta = \Theta^{(D^2\phi)}+\Theta^{(RAQ)}+ \Theta^{(V)}$ in (\ref{eq Theta Eg}) 
we  consider the Newton-Milgrom approximation for a {\em static central symmetric} mass distribution of total mass $m,\;  M=Gm$, in  Euclidean  space (radial distance $r$). Then the Newton potential of the baryonic source is $\Phi^{(bar)}=-\frac{M}{r}$, while the flat space Milrgrom equation (\ref{eq Milgrom equation}) is solved by
 \beq \sigma = \sqrt{a_0 M}\,  \ln \frac{r}{M} \, .   \label{eq central symmetric Milgrom solution}
\eeq 
In spherical spatial coordinates (with $x_1=r$) and the Beltrami-Laplace operator \\ $\nabla^2 = \partial_r^2 + \frac{2}{r}\partial_r$ 
\beq \partial_r \sigma =   \frac{\sqrt{a_0 M} }{r} \, , \qquad \nabla^2 \sigma  =   \frac{\sqrt{a_0 M}}{r^2} \qquad [= - \partial_r^2 \sigma \quad \mbox{(sic)} ].
\eeq
 As  $\sqrt{a_0 M} \ll 1$ for $a_0< a_0$, we find in the MOND region
\beq (\partial_r \sigma )^2 =  \frac{a_0 M }{r^2}  \ll \frac{\sqrt{a_0 M}}{r^2}= \nabla^2 \sigma  \, .
\eeq 
This shows that the entries of $\Theta$ are strongly dominated by the second order derivative terms of $ \Theta^{(D^2\phi)}$. The same holds for the  entries of the other summands of $\Theta$. We  therefore use the {\em approximative} energy momentum tensor $(8\pi G)^{-1}\, \Theta^{\mbox{\em (app)}}$ by reducing  (\ref{eq Theta bras}) to its second order derivative terms, 
\beq \Theta^{\mbox{\em (app)}}_{\mu\nu} 
     \underset{Eg}\doteq   \gamma \left(\, _g\hspace{-0.25em} \nabla_{\hspace{-0.15em}\mu}\partial_{\nu} \sigma  - \frac{1}{2}\,  _g\hspace{-0.2em}\nabla_{\hspace{-0.2em}\lambda}\,\partial^{\lambda} \sigma  \, g_{\mu\nu} 
        \right)\,,  \qquad T^{(\phi)} \underset{Eg}\approx (8\pi G)^{-1}\, \Theta^{\mbox{\em (app)}} \; ,\label{eq Theta approx}
        \eeq
        and work with the approximation $\Theta \underset{Eg}  \approx \Theta^{\mbox{\em (app)}}$  also for the {\em general case} of a Newton-Milgrom approximation.
         In the spatially Euclidean metric this boils down to
\beq  \Theta^{\mbox{\em (app)}}_{\mu\nu} 
     \underset{Eg}\doteq   \gamma \left( \partial_{\mu}\partial_{\nu} \sigma  - \frac{1}{2} \nabla^2\sigma  \, g_{\mu\nu} 
        \right) \, . \label{eq Theta eff}
   \eeq
   This is a peculiar energy-momentum tensor. Its main part has the form of a ``vacuum energy'' tensor with a coefficient ($\nabla^2 \sigma$)  depending, via the Milgrom equation, on the local mass distribution;  but superimposed to it we  also find an additional pressure term $ \partial_{\mu}\partial_{\nu} \sigma $. 
   
For a static or slowly changing scalar field 
we get the result
\beq 
 \Theta^{\mbox{\em (app)}}_{00} \underset{Eg} \doteq   \frac{\gamma}{2} \nabla^2 \sigma\, , \qquad \mbox{ } \qquad   \rho^{\mbox{\em (app)}}  \underset{Eg} \doteq   \frac{\gamma}{16 \pi G} \nabla^2 \sigma \label{eq Theta00 approx}
      \eeq
 and
\beq  tr \, \Theta \approx tr\,\Theta^{\mbox{\em (app)}} \underset{Eg}= - \gamma \, \nabla^2 \sigma \, . \label{eq trace Theta phi}
\eeq
It follows  that the  Newton-Poisson source term (\ref{eq mass density equivalent}) vanishes,
\beq  \rho_N^{(\phi\; \mbox{st})} \; \underset{Eg}\doteq  \quad  0  \, .  \label{eq trace T phi}
\eeq 
 From (\ref{eq Theta bras})  we then read off $\;2 \Theta^{(D^2\phi)}_{00} - tr\,\Theta^{(D^2\phi)} \approx 0$. In the  static case and for $g_{00}\approx-1$ this is true already for the unreduced $\Theta^{(D^2\phi)}$, not only for its approximation.

That shows that the r.h.s contribution of the  scalar field  to the Einstein equation {\em  does not essentially  enter the  Newton approximation} for the Riemannian metrical component in Einstein gauge.
 The only contribution of the scalar field to the acceleration  of freely falling bodies   is contained in the effects of the scale connection (\ref{eq a phi Milgrom approximation}), i.e., 
$  a_{\varphi} = -  \, \nabla \sigma \, .$ 

The formally  calculated mass density which  leads to the same acceleration  in a Newtonian framework is usually called the ``{\em phantom}''  mass density of  modified  gravity. Here it is
 \beq \rho^{(phant)} \underset{Eg} \doteq (4 \pi G)^{-1}\, \nabla^2 \sigma \, . \label{eq phantom matter}
 \eeq  
 For $\gamma=4$ it coincides with the energy density of the scalar field given in (\ref{eq Theta00 approx}) and {\em looses its phantom} character: \label{no phantom character}
 \beq  \rho^{(phant)}  \underset{Eg} \doteq   \rho^{\mbox{\em (app)}} \qquad \mbox{if}\; \gamma=4 \,. \label{eq phantom=app}
 \eeq 
 
\subsection{\small Light deflection due to the scalar field 
 \label{subsection light deflection}}
The scalar field energy-momentum also has crucial repercussions on  gravitational light deflection. We check the spatial components of $h_{jj}$ for a weak field approximation with a static 
 Riemannian metrical component $g \underset{Eg}\doteq \eta + h $  like in (\ref{eq eta + h}).\footnote{For moving matter sources the method of retarded potentials has to be used  \citep[p. 122f.]{Schneider/Ehlers/Falco}.}
   For the central symmetric case we express the Minkowski metric $\eta$ and its perturbation $h$ in spatially spherical coordinates $(x_0,x_1,x_2,x_3)=(x_0, r, \vartheta, \Theta)$, 
\[ \eta= diag\left(-1,\,1,\,r^2,\, r^2  \sin^2 \vartheta \right)\, , \qquad h= diag\left(h_{00},\, h_{11},\, 0 ,\,0 )\right)\, .
\]
This is a special case   of a spherical symmetric metric 
$g=diag(-A,\, B, \, r^2\, r^2 \sin^2 \vartheta) $ with $A= 1-h_{00}, B=1+h_{11}$.
For $\sigma$ we know the the classical solution  (\ref{eq central symmetric Milgrom solution}) of the Milgrom equation,  $\sigma= C \ln r$. As usual we consider the approximation of the Einstein equation solved for the Ricci tensor (\ref{eq Ricci term solved Einstein equ}). 

The first two diagonal components of the  Ricci tensor (Riemannian component only)
 are \citep[p. 123]{Oloff:Raumzeit}
\beqarr R_{00} &=& \frac{A''}{2B}- \frac{A'A'}{4AB}-\frac{A'B'}{4B^2}+\frac{A'}{Br} \nonumber \\
R_{11} &=&\frac{A'B'}{4AB} - \frac{A''}{2A} + \frac{A'A'}{4A^2} +\frac{B'}{Br} \nonumber
\eeqarr
For a weak field approximation   we neglect, as usual, the second degree terms in $h$ and its derivatives. From  (\ref{eq Ricci term solved Einstein equ}), (\ref{eq Theta eff}) we then get:
 \beqarr 
 R_{00} &\approx& - \frac{1}{2} (h_{00}'' + \frac{2}{r}h_{00}') = -\frac{1}{2} \nabla^2 h_{00} \underset{Eg}\doteq   4\pi G\, \rho^{(bar)} \nonumber \\
R_{11} &\approx& \frac{h_{00}''}{2} + \frac{h_{11}'}{r} \underset{Eg}\doteq  - 4\pi G\, \rho^{(bar)} + \gamma \, \partial_r^2 \sigma \label{eq weak field approx light}
\eeqarr 
The first line is  the Newton approximation, $h_{00}= - 2 \Phi_N$. Adding the two equations leads to
\beq \frac{h_{11}'}{r} \underset{Eg}\doteq   \frac{h_{00}'}{r}  + \gamma \, \partial_r^2 \sigma  \label{eq aux light deflection}
\eeq 
For $\sigma=0$ this is the  weak field approximation of Einstein gravity with 
\beq h_{11}=h_{00}= -2 \Phi_N  \, .
\eeq 
But here, for the  spherical symmetric solution (\ref{eq central symmetric Milgrom solution}) of the Milgrom equation,   (\ref{eq aux light deflection})  turns into 
\[ h_{11}' = h_{00}' - \gamma   \frac{\sqrt{a_0 M}}{r}
\] 
and thus  $h_{11}= -2 \Phi_N - \gamma   \sqrt{a_0 M}\ln r$. For 
\beq \gamma =4 \label{eq gamma=4}
\eeq 
this becomes 
\beq
h_{11}=  -2 \Phi_N - 4 \sigma = -2 (\Phi_N+  2 \sigma )
\eeq 
Translated in terms of spacelike Cartesian coordinates that means
\beq h= diag \left(-2\Phi_N,\, 2 (\Phi_N + 2\sigma), \, 2 (\Phi_N + 2\sigma), \,2 (\Phi_N + 2\sigma ) \right)\, \ \label{eq deflection pot central symm}
\eeq 

 In a first order approximation with regard to $h$ and its derivatives, the deflection angle $\alpha$ of the  spatial wave vector of a small wave package travelling along null geodesics can be expressed  in terms of the spatial gradient $\overrightarrow{\nabla}$ of  $\frac{1}{2}(-h_{00}+ h_{jj})$ 
 for any $1\leq j \leq 3$, assuming that all three spatial $h$ are equal \citep[p. 288f.]{Carroll:Spacetime}. In the literature often only a pressure-free matter source is discussed. Then   $h_{00}=-h_{jj}$, and the deflection potential is simply $-h_{00}=2\Phi_N$  In our case 
with  $h_{00}\neq h_{jj}$  the {\em deflection potential}   is (see appendix \ref{Appendix light deflection}):
\beq
\frac{1}{2}(-h_{00}+h_{jj}) = 2(\Phi_N+\sigma) \label{eq lensing potential}
\eeq 
With this  approximation we arrive at the \\[-0.8em]

{\bf Result}: a) In the Newton-Milgrom weak field approximation of a spherically symmetric baryonic matter distribution, the {\em invariant potential} $\sigma$ of the scalar field  {\em contributes to gravitational light deflection} via the $h_{jj}, \, 1\leq j \leq 3,$ in addition to  the Newton potential $\Phi_N$ which is   due only to baryonic matter (both taken twice).\\[-1em]

b) For $\gamma=4$ the effects of the scalar field for  light deflection  are exactly those of an  enhancement of the Newton potential by $\sigma$. This is in  quantitative agreement with the additional acceleration induced by the scale connection $a_{\varphi}$ in (\ref{eq a phi Milgrom approximation}), if one emulates the latter by a formal enhancement of the Newton potential $\Phi_N$ . \label{result b} \\[-0.8em] 

If astronomical observations should indicate a significant difference between the gravitational potentials for  trajectories and for light deflection of simple systems in the dark matter sector, our model could accommodate this by lowering or raising $\gamma$; but at the moment there is no reason to do so.\footnote{For the relation of virial mass and lensing mass for clusters see sec. \ref{subsection clusters}.} 

The above result  for the central symmetric case suggests the {\em conjecture} that also for the general case we  can reasonably expect a similar close relation between the potentials for light deflection   and  for gravitational acceleration, induced by a scalar field with Lagrangian (\ref{eq LDphi}) including the term (\ref{eq L-bras}).

\section{\small A relativistic generalization of MOND  \label{section MOND}}

\subsection{\small  Comparison with MOND for simple systems \label{subsection MOND}}
A look at (\ref{eq central symmetric Milgrom solution}) and (\ref{eq a phi Milgrom approximation}) shows that for the central symmetric case the  additional acceleration of the scalar field in the Milgrom approximation agrees with the  deep MOND acceleration of the classical MOND algorithm. For a comparison between the latter  and the present model we introduce the following\\[0.3em]
\noindent
{\bf Terminology}: In a Newton approximated relativistic weak field regime with baryonic Newton acceleration $a^{(bar)}$ we distinguish between the following regions (with, e.g., $k=l=1$):\footnote{$l$ is to be chosen such that$10^{-2l}\, a_0$ is smaller than the observational errors for accelerations; then the total acceleration of our model in the deep MOND region  is well approximated by  the additional acceleration of the scalar field  $a_{\varphi}$.}
\begin{tabbing}
\hspace*{0.54em} \= (i) the {\em Newton region} for \hspace{1em}   \=   $ 10^{2(k+l)} \, a_0\;$\= $\leq \;|a^{(bar)}|$   \\
\> (ii) the   {\em  intermediate region}  \> \hspace{1em} $10^{2k}\, a_0  $ \> $\leq \; |a^{(bar)}| \leq  \; 10^{(2(k+l)}\, a_0 $  \\
\> (iii) the {\em MOND region} \> \> \hspace{0.9em}  $ |a^{(bar)}|  \leq\;   10^{2k}\, a_0  \quad (\; \leftrightarrow a_{\varphi}=|\nabla\sigma|\leq 10^k a_0)$ \\
\> \qquad containing the  {\em deep MOND}  region  \> \> \hspace{0.9 em}   $|a^{(bar)}| \leq \;  10^{-2l}\, a_0  \quad (\; \leftrightarrow a_{\varphi}\leq 10^{-l} a_0)$ \\[-1.5em]
\end{tabbing} \label{terminology}
The MG regime of our model covers the MOND region (iii); the Newton region (i) falls into the Newton approximated eEG regime, the intermediate regime (ii) is  described by the model only formally (eq. (\ref{eq intermediate region})).
 
 In the central symmetric case (\ref{eq central symmetric Milgrom solution})  $a_{\varphi} \doteq  \sqrt{a_0 M}r^{-1}$. 
  For a   simple, central symmetric system which is not part of a hierarchical gravitational structure (see below)
 the  present model predicts  gravitational effects on low velocity {\em free fall trajectories   in agreement with MOND dynamics in the deep MOND region}.   With regard to  {\em light bending it differs} like Einstein from Newtonian gravity by the factor 2 from non-relativistic MOND calculations and from RAQUAL (see last section). 

In the   MOND region , i.e. in region (iii),  our model can be characterized by the MOND-typical {\em interpolation functions} $\mu(x)$ and $\nu(x)$ \cite[eq. (8), (10)]{Famaey/McGaugh:MOND}
\beq
a_N = \mu(\frac{a}{a_o}) \, a \; , \qquad \mbox{with} \quad \mu(x)\longrightarrow  \left\{{ 1 \quad  \;  \mbox{for}\; x \to \infty} \atop { x \quad \; \; \mbox{for} \; x\to 0 \; ,} \right. \label{eq mu function}
\eeq 
or the other way round
\beq
a = \nu (\frac{a_N}{a_o}) \, a_N \; ,  \qquad \mbox{with} \quad \nu(y) \longrightarrow 
\left\{{ 1 \hspace{2em} \mbox{for} \; y \to \infty} 
\atop {  y^{-\frac{1}{2}} \; \;   \mbox{for} \; y\to 0 \; . } \right. \label{eq nu function}
\eeq  
Here $a_N$ stands for the Newton acceleration of the baryonic mass.\footnote{$\mu(x)\to x$ means $\mu(x)-x = \mathcal{O}(x)$, i.e. $\frac{\mu(x)-x}{x}$ remains bounded for $x \to 0$.}

For the central symmetric case  our acceleration $a=a_{tot}$ in (\ref{eq a-tot}) with vanishing Newton-Poisson term of the scalar field (\ref{eq trace T phi}) is specified by the interpolation functions 
\beq \mu_w(x) = 1+ \frac{1-\sqrt{1+4x}}{2x} \qquad \mbox{and}
\quad 
\nu_w(y) = 1+ y^{-\frac{1}{2}} \; . \label{eq our mu}
\eeq

Straight forward calculation  (in the approximating Euclidean space) shows that, {\em independent of  symmetry conditions},  
a solution of (\ref{eq Milgrom equation}) is given by $\sigma$ with a gradient  $\nabla \sigma = - a_{\varphi}$ such that 
\beq   a_{\varphi} = \left( \frac{a_0}{|a_N^{(bar)}|}\right)^{\frac{1}{2}}\; a_N^{(bar)}
=     \left( a_0|a_N^{(bar)}|\right)^{\frac{1}{2}}  \; \frac{a_N^{(bar)}} {| a_N^{(bar)}|}   \,  .\label{eq solution non-linear Poisson equ}
\eeq
The solution of the Milgrom equation (\ref{eq Milgrom equation}) is  simpler  than one might expect.  In a first step the linear Poisson equation of the Newton theory is to be solved (\ref{eq baryonic Poisson equation}), then an algebraic transformation of type (\ref{eq solution non-linear Poisson equ}) leads to the acceleration given by  the non-linear Poisson equation (\ref{eq Milgrom equation}). 
In this sense our MOND approximation can be solved by means of a quasilinear  procedure similar to the so-called ``QMOND'' approach in the literature \citep[eq. (30)]{Famaey/McGaugh:MOND}. 
 
If our model is realistic the scalar field's energy density and pressure represent properties of a real entity which expresses its  mark on the {\em gravitational light deflection}. This is a {\em crucial difference} to the classical MOND algorithm and to RAQUAL.

 Another important difference results from the following observation: In hierarchical structured gravitating systems, in particular those allowing for Newton-Milgrom approximations at different scales,  the scalar field energy $\rho^{(\phi)}$ and the energy-momentum tensors $T^{(\phi)}$ of different scales can superimpose. Therefore {\em the scalar field energy density  contributions of   small scale systems  have  to be taken into account in the weak field approximation at a higher level.} This is different  for the pressure terms. In contrast to energy and momentum and the contributions to the scale connection,  they are not additive with volume aggregation in a common reference frame and do not allow to form  mean values at large scales so easily  from their values at small scales.

The other way round, the barycenter of a sufficiently strongly bound subsystem $\mathfrak{A}$ of a larger gravitating  system $\mathfrak{B}$ falls freely in the gravitational field of the latter. If for  a Newton approximation of  $\mathfrak{A}$  the gravitational influences of  $\mathfrak{B}$ including its tidal effects can  be neglected,  the gravitational effects of $\mathfrak{B}$ do not enter the Newton-Milgrom approximation of $\mathfrak{A}$ and can abstracted from, as long as regions with total acceleration (\ref{eq baryonic Poisson equation}) above the acceleration of  $\mathfrak{B}$ are concerned.  This does not hold for  gravitationally weakly bound subsystems and the outer regions of strongly bound subsystems. Here already the Newton approximation (\ref{eq baryonic Poisson equation}) is precarious and the Milgrom approximation (\ref{eq Milgrom equation}) for the scalar field is ill defined. In the present model this seems to be the reason for  the  {\em external field effect} (EFE) of MOND, which  has been observed in the dynamics of weakly bound gravitational subsystems in MOND theory \citep[sec. 6.3, 6.5.2]{Famaey/McGaugh:MOND}.

\subsection{\small A short look at galactic dynamics \label{subsection galaxies}}
\begin{figure}
\center \includegraphics[scale=1]{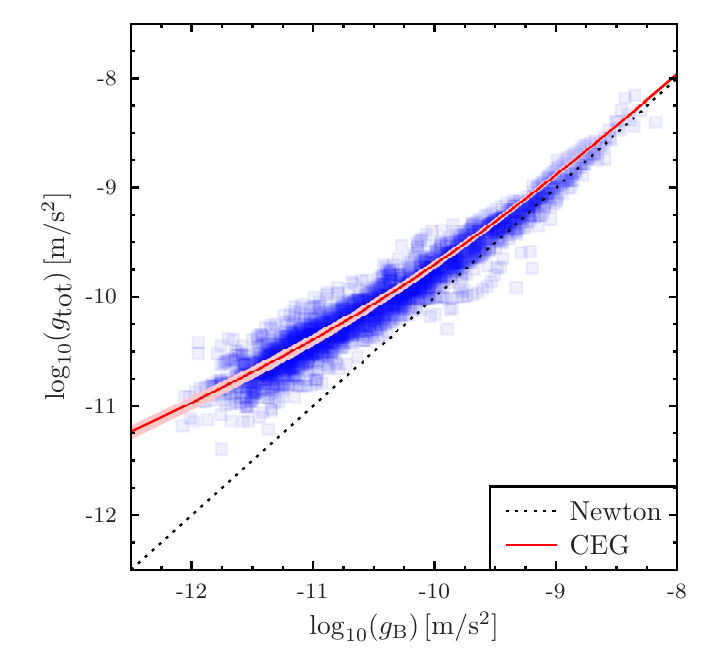} 
\caption{\small Blue squares: Observed, total acceleration
$g_{tot}$ compared with acceleration $g_B$ due to baryonic mass  for 2693 measured data points of 153 galaxies  \citep{McGaugh_ea:2016}.  Red, solid curve  {\em CEG} = graph of
 $g_{tot}$ calculated with $\mu_w(g_B)$  (\ref{eq our mu}), according to our notation. Pink shading indicates $1\,\sigma$ uncertainty.
 Dashed, black line: Newtonian gravity without dark matter. Evaluation and figure due to  \citep{Hossenfelder/Mistele:2018}. \label{fig Hossenfelder/Mistele}} 
 \end{figure}

Based on the hypothesis of a covariantly reformulated version of Verlinde's ``emergent gravity'' (CEG) proposed in \citep{Hossenfelder:2017},
 S. Hossenfelder and T. Mistele  study  a model for galaxies which leads to  MOND-like dynamics  with 
 the same interpolation functions as our (\ref{eq our mu})   \citep{Hossenfelder/Mistele:2018}.
  They   evaluated the data of  2693 measurements referring to 153 galaxies, documented in  \citep{McGaugh_ea:2016}, and 
find  an excellent fit for the radial-acceleration relation of galaxy rotation curves. The empirical check depends only on the interpolation function  (\ref{eq our mu})  and therefore applies just as well to the present model (fig. \ref{fig Hossenfelder/Mistele}).\footnote{One has to keep in mind that our $\mu_w, \nu_w$ are reliable  in the MOND  region (iii) only; they do not apply to  intermediate region or even the eEG regime. }
 
  In a recent paper \citep{Hossenfelder/Mistele:2020} the same authors study  the hypothesis of a condensate similar to the one of  Bherezhiani/Khoury and formulate a model which,   in its   superfluid phase,  leads (``in an idealized limit'') to  MOND-like dynamics again   with  the same  interpolation function (\ref{eq our mu}).
 They investigate whether such a model is consistent with empirical data from  \citep{McGaugh:spiral-arms,Olling/Merrifield:2001} on the rotation curve of  the Milky Way. According to their analysis this is the case, if one allows a moderate rescaling of the baryonic mass by a factor $f_b=0.8$ which lies  in the range of the observational uncertainty. They correctly remark that the result applies also to the  approach of covariant emergent gravity (CEG), mentioned above, because  it ``reduces to the same equations'' as the superfluid model (in the  mentioned idealized limit). The same is clearly the case for the present scalar field model.

A crucial difference between the approaches  has to be kept  in mind.   Following Berezhiani/Khoury,              
the authors assume a pressure less dark matter behaviour of the condensate at the level of galaxy clusters, in fact even already at the outer regions of galaxies (beyond $77\, kpc$). As we have seen, this is not the case for our model. Here we have to estimate the aggregation of scalar field energy contributions in the  halos of the Newton-Milgom approximation of galaxies in the cluster and to add the effects to the Newton-Milgrom approximation of the cluster as a whole, i.e. we have to study the model's prediction for hierarchical systems.
\\[0.2em]

\subsection{\small From stars to galaxies and from galaxies to clusters \label{subsection clusters-1}}
 Let us shed a first glance at two cases of hierarchical systems in the light of the foregoing remarks: (i) the build-up  of galaxies ($\mathfrak{B}$) from stars ($\mathfrak{A})$, (ii)  the  composition of galaxy clusters ($\mathfrak{B}$) from  hot gas ($\mathfrak{A}'$) and galaxies ($\mathfrak{A}$).  In these constellations the Newton-Milgrom approximation for each of the small structures $\mathfrak{A}$ (in its respective  barycentric reference system) allows to calculate  approximately its  scalar field halo with energy momentum $T_{\mathfrak{A}}^{(\phi)}$.  After transformation to the barycentric system of the large structure $\mathfrak{B}$, their sum $\sum_{\mathfrak{A}} T_{\mathfrak{A}}^{(\phi)} $  aggregates to a  collective scalar field halo which we denote by $\overline{\sum_{\mathfrak{A}} T_{\mathfrak{A}}^{(\phi)} }$. Moreover,  the Newton-Milgrom approximation of the global structure in the barycentric reference system of  $\mathfrak{B}$  with baryonic mass density $\rho_{\mathfrak{B}}^{(bar)}$ leads to an expression  $T_{\mathfrak{B}}^{(\phi)}$ for the energy momentum of a global scalar field halo of $\mathfrak{B}$. In the case (i) the aggregation can be understood as an averaging procedure over  spacetime regions on an intermediate scale and we may assume that $\overline{\sum_{\mathfrak{A}} T_{\mathfrak{A}}^{(\phi)} } \approx T_{\mathfrak{B}}^{(\phi)}$.  In the case (ii) the situation is more complicated because of the intervening hot gas component $\mathfrak{A}'$.  The relation between the aggregated and the global halos are different in the two cases.
 
 (i) is a well established case in the MOND framework  \citep[sec. 6.4--6.6]{Famaey/McGaugh:MOND}. In principle it  can be treated  similarly in the present model. Usually a continuity model for the baryonic mass distribution $\rho_{gal}^{(bar)}$ of a galaxy is used. It is  realistic for the gas (neutral hydrogen) inside galaxies, but  an idealization for the stellar masses. The   MOND acceleration at the galaxy level $a_{M gal}$ can be calculated from the  Newton acceleration $a_{N gal}$ generated from $\rho_{gal}^{(bar)}$ by applying the $\nu$-function of (\ref{eq nu function}) (or the QMOND procedure) \citep[p. 58f.]{Famaey/McGaugh:MOND}. 
 From our point of view, the superposition of the scalar field halos (\ref{eq Theta eff})
 of the single stars $\sum_{\ast}  T_{\ast}^{(\phi)}$ reappears  in this procedure aggregated  in the form of a continuity  representation of the scalar field halo of the galaxy as a whole 
 $T_{gal}^{(\phi)}$. It is (calculated in the Newton-Milgrom approximation in the galacto-barycentric system). This procedure presupposes a silent substitution of the spatial average density $\overline{ \sum_{\ast}  T_{\ast}^{(\phi)}}$ of the scalar field halos of the stars $ \sum_{\ast}  T_{\ast}^{(\phi)}$ by the global model for the galactic halo $T_{gal}^{(\phi)}$. 
 The acceleration effects of the scalar field are given by the scale connection at the galaxy level $\varphi_{gal}$  and the corresponding acceleration $a_{\varphi\, gal}$.
 In the usual MOND approach the latter can be expressed by  a  {\em phantom mass density} formally associated to  the  MOND-acceleration field $a_{M gal}$. 
  In our approach this is different.  The aggregated scalar field halos of the stars and the overall scalar field halo of the galaxy seem to be  more or less equal,   $\overline{\sum_{\ast} T_{\ast}^{(\phi)}} \approx T_{gal}^{(\phi)}$, with $T_{gal\, 00}^{(\phi)}= \rho_{gal}^{(\phi)}$ the energy density of the galaxy's scalar field halo in the galacto-barycentric reference system. In our model $\rho_{gal}^{(\phi)}$ expresses a veritable energy density (\ref{eq phantom=app}); it is {\em no phantom}.
 \\[0.5em] 
Two questions have to be posed here:\footnote{Of course there are more questions to ask; for example:   Can relativistic corrections to the external field effect for the solar system in the Milky Way galaxy be estimated and tested (cf. \citep[p. 57]{Famaey/McGaugh:MOND})?
 }

  (i-a) Is the aggregation procedure  $\overline{\sum_{\ast}  T_{\ast}^{(\phi)}} \approx T_{gal}^{(\phi)}$  consistent? 
 
 (i-b) Is the result  compatible with empirical evidence? This is more or less the same as asking whether our interpolation functions  (\ref{eq our mu}) are empirically acceptable at the level of galaxies. \\[-0.4em]

  \noindent
Reasons to be optimistic with regard to the second question are given in a comparable constellation in \citep[sec. 5.3]{Scholz:MONDlike}, but a detailed answer can only be given by astronomers. The first question is of a more theoretical nature, but a positive answer to  the  second one might be taken as an empirical indicator for a positive answer to the first one. \\

Case (ii) is more subtle. Even if we leave  aside the difficulties in gaining realistic mass density profiles for the hot gas in clusters $\rho_{clust}^{(gas)}$ and a continuity model for the star mass density profile $\rho_{clust}^{(\ast)}$, both  adding up to the baryonic mass density  $\rho_{clust}^{(bar)} = \rho_{clust}^{(gas)}+ \rho_{clust}^{(\ast)}$ of the cluster\footnote{See, e.g., \citep{Reiprich/Zhang_ea:2011,Reiprich/Zhang_ea:Corr,Reiprich:Diss}}, there arise {\em  new  problems} at the cluster level. 
In the usual calculation  of the MOND acceleration  $a_{M clust}$ from the baryonic density $\rho_{clust}^{(bar)}$ via its  Newton acceleration $a_{N clust}$ like in (i), 
the  dominance of the hot gas mass over star/galaxy mass leads to a lower contribution of the star mass to the  MOND acceleration $a_{M clust}$ and the associated phantom mass $\rho_{M clust}^{(phant)}$ at the cluster level than one would expect if the galaxies were considered on their own (without the hot gas). The same holds for the additional acceleration $a_{\varphi, clust}$ at the cluster level in our approach -- and here it has consequences. 

Each of the galaxies is a subsystem  $\mathfrak{A}$ freely falling in the gravitational field of the cluster $\mathfrak{B}$ and has a scalar field halo which can be calculated in the Newton-Milgrom approximation of the corresponding galactocentric reference system. The totality of them superposes to $\sum_{gal} T_{gal}^{(\phi)}$. If we want to estimate its averaged value   $\overline{\sum_{gal} T_{gal}^{(\phi)}}$ over  intermediate distance scales (between galactic scale and cluster scale) in the  continuity model, we have to consider  the idealized global mass distribution  $\rho_{clust}^{(\ast)}$ of the galaxies and calculate the corresponding Newton-Milgrom approximation at the cluster level  {\em separately from the hot gas}. This leads to a   scalar field halo of the stars at the cluster level $T_{clust}^{(\phi\, \ast)}$ and a scale connection $\varphi_{\ast}$ (in the cluster barycentric reference system and in Einstein gauge) with the corresponding acceleration  denoted by $a_{\varphi\, \ast}$. If we are optimistic, we may assume  that the continuity calculation gives an estimate for the averaged scalar halos of all the galaxies like in the case (i), $\overline{\sum_{gal} T_{gal}^{(\phi)}} \approx T_{clust}^{(\phi\, \ast)}$. Similarly the scalar field halo of the  hot cluster gas $T_{clust}^{(\phi \, gas)}$ can be estimated  by the Newton-Milgrom approximation with the baryonic content $\rho_{clust}^{(gas)}$ alone. It  contributes   a term $\varphi_{_{gas}}$ to the scale connection and induces the additional acceleration   $a_{\varphi\, gas}$ according to (\ref{eq add acc 1}). The accelerations  $a_{\varphi\, \ast}$ and $a_{\varphi\, gas}$ are due to the scale connection. They are not phantom and their effects add up. They are closely related to the corresponding (approximate) scalar field energy densities $\rho_{clust}^{(\phi\, \ast )}$ and  $\rho_{clust}^{(\phi \, gas )}$ calculated according to eqs. (\ref{eq Theta00 approx}), (\ref{eq phantom matter}), (\ref{eq phantom=app}). Of course, the Newton approximation at the cluster level leads to a Newtonian acceleration $a_{N\, clust}$ sourced by the total cluster mass density $\rho_{clust}^{(bar)}$. The total acceleration in the cluster model is then
\beq a_{clust\,tot} = a_{N\, clust}+ a_{\varphi\, gas} + a_{\varphi\, \ast}
\eeq 
  
If one wanted to emulate the same acceleration in a Newtonian model, the  eqs. (\ref{eq phantom matter}) and (\ref{eq phantom=app}) show that
 one has to take account of the combined mass density of baryonic origin and the two  constitutive parts of the scalar field halo;:
\[ \rho_{clust}^{tot}= \rho_{clust}^{(bar)} + \rho_{clust}^{(\phi \, gas )} + \rho_{clust}^{(\phi\, \ast )}
\]

The lensing of  clusters arises partially from the weak field approximation of the global cluster model with the hot gas as baryonic source which incorporate effects from the cluster scalar field halo in addition to  the Newton potential. But  the total {\em lensing mass} of the cluster incorporates also the contributions due to the micro-lensing effects of all the halos of the galaxies. Here  {\em the Newton-Milgrom approximation at the cluster level has to be complemented by an  estimate of the aggregated effects of the galactic scalar field halos}, which are not taken into  account for the global continuity model with $\rho_{clust}^{bar}$ as baryonic source because of the dominance of the hot gas component in the latter. In these respects the {\em present model differs considerably} from the original MOND estimates at cluster level.

\subsection{\small A heuristic discussion of cluster dynamics \label{subsection clusters-2}}
If the above optimistic approach for estimating the averaged scalar field halo of the galaxies is justified, our model may be able to  explain  cluster dynamics without additional dark matter.
     A  heuristic check  with observational data of 17+2 clusters 
    taken   \citep{Reiprich/Zhang_ea:2011,Reiprich/Zhang_ea:Corr},  evaluated like in   \citep{Scholz:Clusters} for a similar, although in its justification problematic,  model  gave encouraging  results. For 15 of the 17 outlier reduced clusters the observational errors and the model spread overlap (for the remaining two a minor extension leads to overlapping). 
     Among the 15 is  the famous Coma cluster with an  observational value for the total mass 
      $M_{500}= 6.55^{\pm 2.36}$  $10^{14}M_{\astrosun}$ inside the reference distance   $r_{500}=1278 \, kpc$.\footnote{$r_{500}$ is  the   
 distance from the cluster center, at which the total mass density, reconstructed in the framework of a dark matter paradigm from observational data, has fallen down to $500\,\rho_{crit}$, with  $\rho_{crit}$ the critical density of cosmology.  $M_{500}$ is the values for the total mass inside the respective reference distance, determined  from observational data in the framework of the dark matter paradigm. }
 The total mass equivalent of   $\rho_{clust}^{tot}$,  calculated in our model including the scalar field halos of the hot gas and the galaxies and integrated up to the reference distance, is  $M_{tot}(500) \approx 5.7^{+0.9 \atop -0.7} \cdot  10^{14}M_{\astrosun}$. That  compareswell with the observational value  and leaves no mass gap like in the usual MOND approach, see figure 1. 
 \begin{figure}
\center{\hspace*{0em} \includegraphics*[scale=1.2]{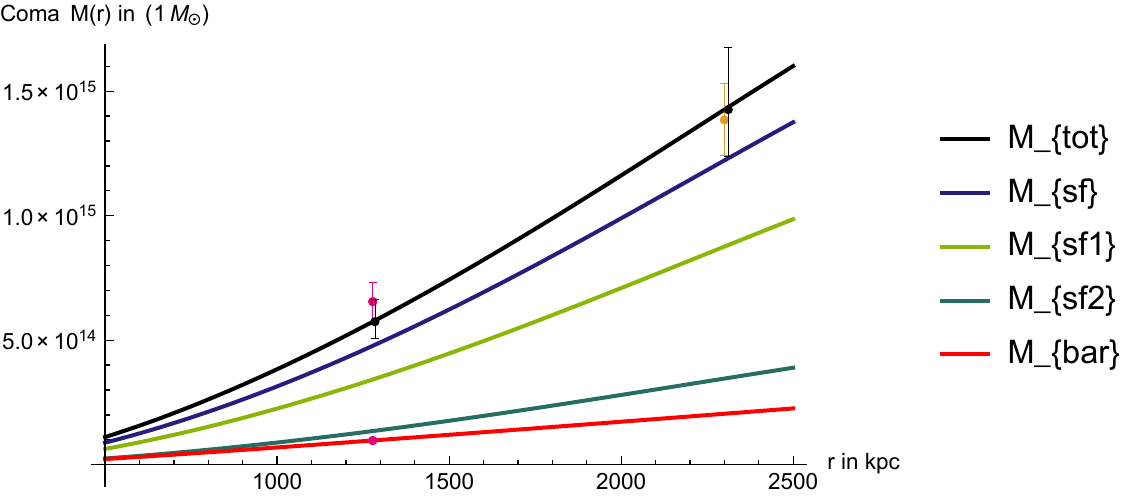}}
\caption{\small Development of mass contributions  in the  model for the Coma cluster, depending on the distance $r$  from the center: Baryonic  mass (hot gas and galaxies) $M_{bar}$ (red) interpolated by a $\beta$-model (see \citep{Scholz:Clusters}), contribution of the scalar field halo $M_{s\hspace{-0.1em}f  2}$ of the galaxies (dark green),  the hot gas $M_{s\hspace{-0.1em}f  1}$ (bright green), the combined scalar field halo $M_{s\hspace{-0.1em}f} = M_{s\hspace{-0.1em}f  1}+ M_{s\hspace{-0.1em}f  2}$ (blue), and the total mass 
 $M_{tot}= M_{s\hspace{-0.1em}f }+M_{bar}$ (black). 
Observational values for the baryonic mass (red dot) and the virial mass  at $r_{500}= 1280 \, kpc$ (red error interval), the latter also  at $r_{200}= 2300 \, kpc$ (orange error interval). Black error intervals at $r_{500}, r_{200}$ indicate model spread resulting from the variation of observational input data.}
\label{fig halo Coma}
\end{figure}
  This may be  a motivation for more detailed, not only heuristic investigations.\footnote{By several reasons the calculations of \citep{Scholz:Clusters,Scholz:Clusters_corrigendum} are problematic; see appendix \ref{Appendix Flaws}. But the numerical results do not differ strongly from the present ones based on a reliable derivation. For example the value of the total mass equivalent of the Coma cluster   
 given in \cite[table 4]{Scholz:Clusters_corrigendum} is    $M_{tot}(500) \approx 5.7^{+1 \atop -0.7} \cdot  10^{14}M_{\astrosun}$.}
 \\[-0.5em]
 
   \noindent
   For a  better understanding of cluster dynamics   questions similar to  (i-a,i-b) have  to be tackled:
   
   (ii-a) Is the above mentioned ( ``optimistic'') estimate of the collective scalar field halos of the galaxies in the continuity model $\overline{\sum_{gal} T_{gal}^{(\phi)}} \approx T_{clust}^{(\phi\, \ast)}$ consistent with, e.g., numerical simulations?  
   
   (ii-b) Are the effects of the superposition of scalar field halos of all the galaxies in the cluster and the halo of the hot gas (of course considered in addition to the Newton potential) generated by the total  baryonic mass sufficient to explain the observational data of clusters?\\
 
Aside from these questions, let us shed a short glance at the bullet cluster 1E0657-56. It is often interpreted as  providing direct evidence in favour of  particle dark matter and of ruling out  alternative gravity approaches. This argument does not apply to the present model. The energy content of the scalar field halos of the  colliding clusters  endows them with  inertia  of their own. The shock of the colliding gas exerts dynamical forces on the gas masses only, not directly on the scalar field halos. We can expect that, during the encounter, the halos will   roughly  follow the inertial trajectories of their respective clusters before collision, and they will continue to do so for a while. Only after a certain time delay a re-adaptation of the mass systems and the respective scalar field halos can take place. Clearly the MOND-approximation of the present model is unable to cover such violent dynamical processes. It  describes only the relatively stable states before collision and -- in some distant future -- after collision. But a separation of halos and gas masses for a (cosmically ``short'') period is to be expected, just like in the case of a particle halo with appropriate clustering properties. 

For the time being, the cluster  1E0657-56 does not help to decide between the overarching alternative  research strategies, particle dark matter or alternative gravity; all the more so for modification by a  scalar field carrying a non-negligible amount of energy-momentum like in the present approach. It may be able to do so, once  the dynamics of gas and of the halos has  been modelled  with sufficient precision in both approaches. Only then a proper comparison can be made; but that is an overtly complicated task. It  seems more likely that other types of observational evidence will offer a simpler path to a differential evaluation of the two strategies  and  help  clarifying  the  alternative.  \\[-0.5em]

 Here, even more than in the case of the galaxy dynamics, a reliable judgement on the empirical feasibility  of the model can only be given by astronomers. For the moment we have to content ourselves with the theoretically intriguing properties of the energy momentum tensor of the scalar field, which may indicate a new route towards solving the missing mass problem for clusters.

\section{\small Discussion and outlook \label{section Discussion}}
The framework of the model studied in this paper can be  viewed as a kind of generalized  Jordan-Brans-Dicke approach, but it becomes more transparent if it is formulated in terms of integrable Weyl geometry. It leads to a  modification of Einstein gravity  with a scale invariant Einstein equation   (\ref{eq Einstein equation}). We have assumed that the scalar field appears in  two  phases  governed by different Lagrangians in the eEG regime and the MG regime. In the first one the scalar field is governed by a usual quadratic kinetic Lagrange term with $\alpha \gg 1$; here the dynamics agrees effectively with Einstein gravity. The present paper concentrates on the MG regime in which Einstein gravity is  modified for very very weak gravitational field constellations.  Here the scalar field Lagrangian consists of  a cubic kinetic term typical for MOND-like behaviour (\ref{eq LDPhi3 gen}), a conformally coupled quadratic kinetic term,  and a   second order kinetic self-interaction term important for the self-energy of the scalar field  (\ref{eq L-bras}).

 In the Einstein gauge of the Weylian metric (``Einstein frame'') the gravitational equation of the MG regime can be brought into the form of an  Einstein equation (\ref{eq Einstein eq MG}) for the Riemannian component $g$ of the Weylian metric.  The scalar field $\phi$,  in Riemann gauge (``Jordan frame'')\footnote{See fn. \ref{fn gauge/frame}.} 
 written as $\tilde{\phi}= \phi_0 e^{-\sigma}$, is governed by 
 a covariant version of Milgrom's nonlinear Poisson equation for the exponent $\sigma$ (\ref{eq covariant Milgrom equation}). Accordingly, for  a weak field, quasi-static constellation   one has to consider weak field approximations   for the two  constituents $g$ and $\varphi$ of the Einstein gauged Weylian metric, where $ \varphi = d \sigma$.   The Einstein equation for $g$ leads to  a Newton approximation (\ref{eq Poisson equation}) which is sourced by the baryonic matter and the scalar field. Surprisingly, the additional source term of the scalar field (\ref{eq mass density equivalent}) vanishes for our Lagrangian (\ref{eq trace T phi}). The  scalar field equation for $\sigma$ simplifies to the classical Milgrom equation in Euclidean space (\ref{eq Milgrom equation}). Both together constitute the   {\em Newton-Milgrom} weak field {\em approximation} of the present theory. 

 In this approximation the kinematics of freely falling test bodies  (\ref{equ of motion}) is influenced  by the invariant potential $\sigma$ through a modification of the affine connection in Einstein gauge (\ref{eq add acc 1}). For an appropriate choice of the coefficients 
 (\ref{eq a_0}) this leads to a MOND-like phenomenology, with   interpolating functions  (\ref{eq our mu}) taken into closer considerations already by Hossenfelder/Mistele  based on a different Lagrangian approach.  Of course, the latter are valid only  in the MG regime. In the eEG regime, the influence of the scalar field is negligible; as already said it is  effectively governed by Einstein gravity. In the MG regime the  role of the scalar field  as a source term for the Einstein equation (\ref{eq Einstein eq MG}) and for the weak field  approximation  used for calculating light deflection (\ref{eq weak field approx light})   has  important consequences  for gravitational lensing, which is {\em different from both classical MOND and RAQUAL}. For the central symmetric case it is given by (\ref{eq deflection pot central symm}).
 
 For hierarchical systems like galaxies, formed from stars and interstellar gas,  the scalar field of the global system may be approximated by the averaged scalar field of the stars as constituent parts. For clusters, formed from galaxies and hot gas,  this is different because of the dominance of the hot gas in the total baryonic matter. Here the scalar field halos of the galaxies have to be taken into account  in addition to the  halo of the hot gas. This is clear for the lensing mass of the cluster and can be reasonably  assumed also for its virial mass. This may suffice to close the mass gap arising in the usual MOND approach to cluster dynamics.
  \\[-1.3em]

  \noindent
A series of open questions remains:\\[-1.3em]

On the {\em theoretical level}  the contribution of the scalar field to the light deflection potential  has to be worked out for the general case. Moreover 
it remains to be checked, whether  the propagation velocity of scalar field excitations is  different from RAQUAL because of the second order kinetic   term (\ref{eq L-bras}). And even if not, the difficult question of group velocity versus phase velocity had to be decided, before one comes to a final judgement of the physical feasibility of the approach.

On the {\em empirical side} there are  important questions of  adequacy  at  the scale of galaxies and of galaxy clusters (cosmological questions belong to an epistemologically different class, see below):  (i-b) Can  the  interpolating  functions given in (\ref{eq our mu}), which arise from the additive overlay of the scalar field effects to the baryonic Newton dynamics, reproduce the successes of the classical MOND algorithm for galaxy rotation curves? Encouraging first positive results are given in \citep{Hossenfelder/Mistele:2018,Hossenfelder/Mistele:2020}. \hspace{0.1em}
(ii-b) Are the estimates of the total virial mass and the lensing mass of clusters in agreement with observational data, possibly even without assuming additional dark matter?

Another class of questions refers to a possible {\em material underpinning} for the Lagrangian (presupposing a positive answer to the questions of empirical adequacy). It seems  unlikely that the  scalar field $\phi$ of the  model represents a fundamental field. If it is physical, it is much more likely that it  expresses some collective state comparable to the superfluid approach to DM/MG. But the energy-momentum tensor 
(\ref{eq Theta eff}) shows more similarities with what is usually considered as a ``dark energy'' tensor than with dark matter. The different subcategories of the dark sector seem to be moved closer together  than  is usually thought. 
This may indicate a problem for the possibility of bringing the present scalar field model in closer relation with  the superfluid approach; but  the question remains open.  

From a  different perspective, the biquadratic potential of $\phi$ with the Higgs field (\ref{eq biquadratic potential}) can be used to  establish a (weak) connection between our field and the {\em Higgs portal}. In this case the hierarchy factor $\eta$ in the potential  has to be explained and  the question of the fundamental constituents for the collective state function $\phi$ is posed. A -- very speculative -- possibility is studied in approaches like \citep{Cheng:1988,Ohanian:2016,Ghilencea:2019}, for which  the integrable scale connection, and its scalar field $\phi$, arises as a collective limiting case of a non-integrable scale connection $\varphi$ with a ``true'' Weyl vector boson is being studied. This may also lead to different view of the distinction between JBD and MG regime assumed here: If the scalar field arises as a collective excitation state of a quantum field and vanishes in higher curvature regions of space-time, the scalar field in the eEG regime may be just a formal fiction, and the non-MG regime could turn out to be governed by Einstein gravity per se. In this case the Lagrangian density of the scalar field  in (\ref{eq LDphi}) for the eEG regime would  have to be substituted by  a Lagrange constraint $\sum_{\nu}\lambda_{\nu}D_{\nu}\phi$ for the  {\em Einstein regime}, setting $D_{\nu}\phi=0$ (for all $\nu$) and enforcing the identity of Riemann gauge with the  Einstein gauge \citep{Ghilencea:2019}.

Questions of {\em cosmology} have not been dealt with here. It seems not yet clear, whether the  MG regime of the present model extends to the cosmological scale, and if so which consequences would arise for cosmological models. For studying such questions it should be taken into account that the cosmological problems leading to  DM and of MG have a different epistemic source than the astrophysical ones dealt with here. Cosmological models have an inherent epistemic ambiguity. They try to scientifically represent the  material world as a whole  and thus are always in the danger of over-stretching the scientific method. Although the present standard model of cosmology can   proudly claim many impressive successes, any further going contention of the predominance for cosmological tests over those from astrophysical observations in more ``nearby'' regions (say redshift $z < 3$) indicates  lacking  critical self-reflection of the epistemic status of the cosmological model. It should be rejected or passed over in silence. 

The  recent  Hubble telescope data on the Hubble parameter, which  indicate a $4\, \sigma$ discrepancy between   (cosmologically) model-independent, direct measurements of $H_0$ from the nearby universe and predictions from the ``early'' universe made on the background of the  $\Lambda CDM$-model of cosmology \citep{Riess_ea:2019}, may be a warning sign. The authors of the study draw the conclusion ``A new feature in the dark sector of the Universe appears increasingly necessary to explain the
present difference in views of expansion from the beginning to the present'' \citep[preprint p. 18]{Riess_ea:2019}.
To give this quote here does not mean claiming the status of such a ``new feature of the dark sector'' for  scalar field halos of the present approach. It rather serves  the purpose of underpinning the above mentioned choice of priority for ``nearby'' astrophysical observational evidence over cosmological  criteria which presuppose the $\Lambda CDM$-model. 

In any case, our model also has some interesting features on a  general, so to speak  {\em philosophical level}. It demonstrates that the possibilities for the elaboration of models for the explanation of astrophysical DM phenomena starting from a MG approach in the framework of a classical geometric setting and a simple field content are not yet exhausted. Just to the contrary, our relatively simple scalar field model is based on a  natural, moderate generalization of Riemannian geometry; it  does not need to impute artificially looking structures for the  physical geometry of space-time. Its  kinetic self-interaction term  tends to undermine the presently dominant dichotomy between space-time and the dark sector, a bit like the superfluid approach does with regard to the DM-gravity dichotomy \citep{Martens/Lehmkuhl:2019}. In contrast to the latter the  energy momentum tensor of the scalar field resembles  dark energy more than (dark) matter. We should keep these strange properties of the energy-momentum tensor in mind. They may be a sign that the present approach has a  value as a formal model only; if not, they would seem  to  indicate that a greater shift in the  ontology of the dark sector becomes necessary.\\[5em]


\section{\small Appendix \label{section Appendix}}

\subsection{\small  Weylian metric,  derivative operators, curvatures, Einstein tensor  \label{Appendix IWG}}
There are many introductions to Weyl geometry, among them the classics \citep{Weyl:InfGeo,Weyl:GuE,Weyl:RZMengl,Eddington:Relativity,Pauli:1921/2000,Bergmann:Relativity}.\footnote{Mor recent ones can be found in \citep{Blagojevic:Gravitation,Israelit:1999Book}, \citep[chap. IX]{Tonnelat:1965},  \citep[appendix A]{Drechsler/Tann} and  \citep{Perlick:Diss} (difficult to access). For selected aspects see \citep{Codello_ea:2013} and  \citep[sec. 4]{Ohanian:2016}. Integrable Weyl geometry is presented in \citep{Dahia_ea:2008,Romero_ea:Weyl_frames,Almeida/Pucheu:2014,Quiros:2014a},  \citep[sec. 2.1]{Scholz:2011Annalen}.  Be aware of different conventions for the scale connection.   Expressions for Weyl geometric derivatives  and  curvature quantities are derived in  \citep{Gilkey_ea,Yuan/Huang:2013} and \citep[App.]{Miritzis:2004}. For  a more  mathematical  perspective consult  \citep{Folland:1970,%
Calderbank/Pedersen:1998,Gauduchon:1995,Higa:1993}. \label{fn lit WG} }
Here follows a short introduction  with particular emphasis on the conventions and notations used in this paper.

A Weyl geometric structure can be specified in a scale-gauge dependent manner by a  pair $(g,\varphi)$ of a semi-Riemannian metric $g = g_{\mu\nu}dx^{\mu}dx^{\nu}$, representing the {\em Riemannian component} of the {\em Weylian metric} and a differential 1-form $\varphi = \varphi_{\nu}dx^{\nu}$ representing the {\em scale connection} in the chosen scale-gauge. A change of scale (or gauge) is given by conformal rescaling $\tilde{g} = \Omega^2 g$ accompanied by the gauge transformation $\tilde{\varphi}= \varphi - d \log \Omega$.  The {\em scale invariant affine connection} $\Gamma$ (in coefficients $\Gamma_{\mu \nu}^{\lambda}$) of Weyl geometry is  a sum $\Gamma = \Gamma (g) + \Gamma(\varphi)$ of the Levi-Civita connection $\Gamma(g)$ of $g$ and a part depending on the scale connection, which  in coefficients is given by
\beq  \Gamma(\varphi)_{\mu\nu}^{\lambda}= \varphi_{\mu} \delta_{\nu}^{\lambda}+ \varphi_{\nu} \delta_{\mu}^{\lambda} - g_{\mu \nu}\varphi^{\lambda} \,.  \label{eq varphi Levi-Civita}
\eeq 
To avoid clumsy expression we also use a notation with pre-sub-scripts $g$ and $\varphi$ like $_g\hspace{-0.1em}\Gamma=\Gamma(g)$ and $_{\varphi}\hspace{-0.1em}\Gamma=\Gamma(\varphi)$ etc. $\Gamma$ defines a covariant derivative operator $\nabla = \nabla (\Gamma)$ different from the covariant derivative with regard to the Levi-Civita of the Riemannian component, also  denoted  by pre-sub-script   $_g\hspace{-0.2em}\nabla = \nabla(\Gamma(g))$.

 For  scale covariant fields $X$ of weight $w(X)=w$ (i.e. $X\mapsto \tilde{X}= \Omega^w X$ for a change of scale by $\Omega$ like above) the scale covariant derivative $D$ is given by $DX =  \nabla X + w\, \varphi \otimes X$, e.g. for a vector field $D_{\mu}X^{\nu}= \nabla_{\mu}X^{\nu} + w\, \varphi_{\mu}X^{\nu}$, for a scalar field $D_{\mu} X=\partial_{\mu} X + w \,\varphi_{\mu}X$ etc. It is important to distinguish the three derivations   
 \beq \quad _g\hspace{-0.2em}\nabla X, \qquad \nabla X , \qquad D X \label{eq derivations}
 \eeq
 for scale covariant  fields (scalar, vector or tensor). Lifting and lowering of indices, i.e. transformations between tangent vector components and its duals, are given by the Riemannian component $g$ in a scale gauge and thus change the scale  weight of a field, e.g. $w(X^{\mu}) = w(g^{\mu \nu}X_{\nu})= w(X_{\nu})-2$.
 
 The Weyl geometric  Riemann  tensor is defined as $Riem = Riem (\Gamma)$ and thus scale invariant. The same holds for the Ricci tensor $Ric = Ric(\Gamma)$, while the scalar curvature (Ricci scalar) $R$ uses lifting of indices and is thus of weight $w(R)=-2$. Like for the affine connection one often needs to compose the  Weyl geometric curvature quantities from their Riemannian counter parts (depending on $g$ in one gauge only) and a scale connection part; we write
 \beq Riem = \, _g\hspace{-0.1em}Riem + \, _{\varphi}\hspace{-0.1em}Riem \qquad \mbox{etc.}
 \eeq
The Einstein tensor $G=Riem -\frac{1}{2}R g$ is similarly (de-)composed
\beq G = \, _g\hspace{-0.1em}G + \, _{\varphi}\hspace{-0.1em}G  \, .
\eeq

The curvature of the scale connection $\varphi$  is given by the exterior differential $d\varphi$. If it vanishes ($\partial_{\mu}\varphi_{\nu}-\partial_{\nu}\varphi_{\mu}=0$), the Weylian metric is locally {\em integrable}, i.e., at least for simply connnected regions it can be brought into the form of a Riemannian metric by chosing a gauge with $\tilde{\varphi}=0$. This is called the {\em Riemann gauge}.

\subsection{\small  Some useful formulas \label{Appendix formulas}}
For the Ricci tensor of Weyl geometry $Ric=\, _g\hspace{-0.1em}Ric + \, _{\varphi}\hspace{-0.15em}Ric$, the scalar curvature $R=\, _g\hspace{-0.1em}R + \, _{\varphi}\hspace{-0.15em}R$ and the Einstein tensor $G=\, _g\hspace{-0.1em}G + \, _{\varphi}\hspace{-0.15em}G$ the following relations hold in dimension $n$  for  any scale gauge
\beqarr _{\varphi}\hspace{-0.15em}R_{\mu \nu} &=& (n-2)(\varphi_{\mu}\varphi_{\nu}-\,_g\hspace{-0.2em}\nabla_{\hspace{-0.15em}(\mu}\varphi_{\nu )} ) - \left((n-2)\varphi_{\lambda}\varphi^{\lambda} + _g\hspace{-0.3em}\nabla_{\hspace{-0.15em}(\lambda}\varphi^{\lambda )} \right) g_{\mu \nu} \\
_{\varphi}\hspace{-0.15em}R &=& (n-1)(n-2)\, \varphi_{\lambda}\varphi^{\lambda}- 2(n-1 )\,_g\hspace{-0.2em}\nabla_{\hspace{-0.2em}\lambda}\varphi^{\lambda}  \\
_{\varphi}\hspace{-0.15em}G_{\mu \nu} &=& (n-2) \left(\varphi_{\mu}\varphi_{\nu}-\,_g\hspace{-0.2em}\nabla_{\hspace{-0.15em}(\mu}\varphi_{\nu )}  + ( \frac{n-3}{2}\,\varphi_{\lambda}\varphi^{\lambda} + _g\hspace{-0.3em}\nabla_{\hspace{-0.15em}(\lambda}\varphi^{\lambda )})\, g_{\mu \nu}\right) 
\eeqarr
In our case 
\beq _{\varphi}\hspace{-0.15em}G_{\mu \nu}  \underset{Eg}\doteq 2(\partial_{\mu}\sigma\partial_{\nu}\sigma -\,_g\hspace{-0.2em}\nabla_{\hspace{-0.15em}(\mu}\partial_{\nu )}\sigma ) + \left(\partial_{\lambda}\sigma \partial^{\lambda}\sigma  + 2  _g\hspace{-0.3em}\nabla_{\hspace{-0.15em}(\lambda}\partial^{\lambda }\sigma  \right)  g_{\mu \nu}  \label{eq varphi G}
\eeq 

For $\tilde{\phi} \underset{Rg}\doteq \phi_0 e^{-\sigma}$ the following holds:
\beqarr D_{\mu}\phi &=& \partial_{\mu}\phi- \varphi_{\mu}\phi  
            \underset{Eg}\doteq   -\varphi_{\mu} \phi_0 
           \;  \underset{Eg}\doteq \; -\phi_0 \partial_{\mu}\sigma  \\
           D_{\lambda}(|D\phi|D^{\lambda}\phi) & \underset{Eg}\doteq & - \phi_0 D_{\lambda}(|D\phi|\partial^{\lambda}\sigma)  \nonumber \\
  \mbox{And because of}  \qquad  _{\varphi} \hspace{-0.1em}\Gamma_{\mu \nu}^{\mu} &=& 4 \varphi
           _{\nu} \quad \nonumber \\
      D_{\mu}D^{\mu}\phi &=&  \nabla_{\mu} D^{\mu}\phi - 3 \varphi_{\mu}\,D^{\mu}\phi = \, _g\hspace{-0.15em}\nabla_{\mu} D^{\mu}\phi + \varphi_{\mu} D^{\mu}\phi  \nonumber \\ 
       &\underset{Rg}\doteq & - \tilde{\phi}\, (\, _g\hspace{-0.2em}\nabla_{\hspace{-0.2em}\mu}\partial^{\mu} \sigma - \partial_{\mu}\sigma\partial^{\mu}\sigma )  \nonumber \\
        &\underset{Eg}\doteq & -\phi_0\, (\, _g\hspace{-0.2em}\nabla_{\hspace{-0.2em}\mu}\partial^{\mu} \sigma + \partial_{\mu}\sigma\partial^{\mu}\sigma )
 \eeqarr 
  As $|\nabla \sigma|= \epsilon_{\sigma} \partial_{\sigma}\partial^{\sigma}$  and therefore $w(|\nabla \sigma|)=-2$,  we get
 \beqarr D_{\lambda}(|\nabla \sigma| \partial^{\lambda}\sigma) &=&\, _g\hspace{-0.15em} \nabla_{\hspace{-0.15em}\lambda}(|\nabla \sigma| \partial^{\lambda}\sigma) + \, _g\hspace{-0.15em}\Gamma_{\lambda \nu}^{\lambda}|\nabla \sigma| \partial^{\nu}\sigma - 3 \varphi_{\lambda}\, |\nabla \sigma| \partial^{\lambda}\sigma \nonumber \\
 &=& \, _g\hspace{-0.15em} \nabla_{\hspace{-0.15em}\lambda}(|\nabla \sigma| \partial^{\lambda}\sigma) +  \varphi_{\lambda}\, |\nabla \sigma| \partial^{\lambda}\sigma 
 \label{cubic cancelling} \eeqarr 
    
    For the variation  $\frac{\delta L_{D\phi^3}}{\delta \phi}$ the following modules are helpful:   
\beqarr 
\frac{\partial L_{D\phi^3}}{\partial \phi} &=& - 2 \phi^{-1}\, L_{D\phi^3}, \qquad
\frac{\partial L_{D\phi^3}}{\partial(D_{\lambda}\phi)} = - 2 \beta \xi^3 \phi^{-2}\, |D\phi|D^{\lambda}\phi \nonumber \\
D_{\lambda}\frac{\partial L_{D\phi^3}}{\partial(D_{\lambda}\phi)} &=& -2 \beta \xi^3 \phi^{-2}\, D_{\lambda}( |D\phi|D^{\lambda}\phi  ) +6 \phi^{-1}L_{D\phi^3} \label{eq variation LDphi3}
\eeqarr

\subsection{\small  Geroch-Jang theorem \label{Appendix Geroch-Jang}}
In Einstein gravity (EG) the principle of geodesic motion of test particles is supported by the following theorem of  Geroch and  Jang \citep{Geroch/Jang:1975}.\\[0.5em]
{\em Theorem Geroch-Jang}: \\
After appropriate re-parametrization, a smoothly embedded timelike curve $\gamma$ in an oriented Lorentzian manifold $(M,g)$ is a geodesic if the following holds:$\quad$
For any open neighbourhood $U$ of $\gamma$ (more precisely its image/trace) there is a  smooth symmetric 2-form $T_{ab}$ with support in $U$, which does not vanish identically, is covariantly conserved, 
and satisfies the strong dominant energy condition in the following sense: For every timelike covector field $\xi_a$ the vector field $T^{ab}\xi_a$ is timelike at any point and  $T^{ab}\xi_a \xi_b\geq 0 $.

This theorem may reasonably be interpreted as implying the {\em geodesic principle}: Sufficently small test bodies move along timelike geodesics. 
It is easily imported into the integrable Weyl geometric (IWG) framework by the following argument \citep{Lehmkuhl/Scholz:forthcoming}.\\[0.5em]
{\em Theorem Geroch-Jang (IWG)}: \\
Let $\gamma: I\longrightarrow M$ be a smoothly embedded timelike curve in an oriented {\em integrable Weylian manifold} $(M,[(g,\varphi)])$ satisfying the following properties:
For any open neighbourhood $U$ of $\gamma$ (more precisely its image/trace) there is a  smooth, not identically vanishing, symmetric {\em scale covariant} 2-form $T_{ab}$ of  weight $w(T_{ab})=-2$ with support in $U$, which satisfies the conditions (derivation operators $\nabla$ and $D$ as in (\ref{eq derivations}):\vspace{-0.5em}
\begin{itemize}
\item[(i)] $T$ is {\em scale covariantly conserved}, $D_a T^{ab}=0$.\vspace{-0.5em}
\item[(ii)] $T$  satisfies the strong dominant energy condition in any gauge.\footnote{For every timelike covector field $\xi_a$  $T^{ab}\xi_a$ is timelike  and  $T^{ab}\xi_a \xi_b\geq 0 $ in any scale gauge ($T$ not everywhere zero, here already clear).} \vspace{-0.5em}
\end{itemize}
Then the curve  can  be re-parametrized, $\tilde{\gamma}: I \longrightarrow M$, as  a Weyl geometric geodesic.\footnote{Whether a  scale invariant geodesic or  a  scale covariant geodesic of weight -1,  depends on the re-parametrization.}\\[0.3em]
The proof is easy. Obviously the conditions (i) and (ii) hold in any scale gauge, if they are satisfied in one. Go to Riemann gauge. Then the conditions of the Riemannian Geroch-Jang theorem are satisfied; thus $\gamma$ can be re-parametrized as a Levi-Civita geodesic in Riemann gauge $(\tilde{g},0)$. In $(\tilde{g},0)$ the Weyl geometric derivative $\nabla$ coincides with $_{\tilde{g}\hspace{-0.2em}}\nabla$ (the Levi-Civita derivation of $\tilde{g}$), thus $\tilde{\gamma}$ is a scale invariant geodesic of the Weyl structure (i.e., $\nabla_{\dot{\tilde{\gamma}}}\dot{\tilde{\gamma}}=0$). An appropriate re-parametrization  gives it unit norm in any other gauge.\\[0.1em]

{\em Dynamical interpretation}: If a scale co/in-variant theory of gravity is formulated in  the framework of  {\em integrable} Weyl geometry (IWG), the energy-momentum 2-form of matter  $T_{ab}$ is of weight -2.\footnote{In a Lagrangian formulation the scale invariance of the matter Lagrangian $\mathfrak{L}_m= L_m\sqrt{|g|}$ demands/implies $w(L_m)=-4$. The variational derivative  $T_{ab} =\frac{ \delta L_m}{\delta g^{ab}}$ increases the weight by 2.} Independent of which scale gauge expresses the observable quantities  most directly 
-- in Weyl geometric scalar tensor theory (WST) it is the Einstein gauge  -- a test body may be understood as  the limit of small energy-momentum tubes  in this gauge. Then the conditions (i), (ii) are satisfied in this gauge, therefore  in any one, and the theorem can be applied. 
{\em Result}: The geodesic principle holds in any reasonable  dynamical theory of gravity in  IWG. 

\subsection{\small  Short comment on gravitational light deflection \label{Appendix light deflection}}
According to a method used in \citep[p. 288ff.]{Carroll:Spacetime}   (following  an approach outlined in \citep{Pyne/Birkinshaw:1996}) the deflection angle $\alpha$ of the  spatial wave vector of a small wave package travelling along null geodesics can be expressed, in a first order approximation in $h$,  in terms of the spatial gradient $\overrightarrow{\nabla}$ of  $\frac{1}{2}(-h_{00}+ h_{jj})$ 
 for any $1\leq j \leq 3$ (no summation over $j$). But here, like in other places of the literature, the special case of a pressure-less matter energy-momentum  is assumed. 
A  similar,  in  the result a bit  more general, derivation of the deflection potential (including the case of moving masses) is given in \citep[(4.19), p. 124]{Schneider/Ehlers/Falco}. Here    the authors add an explanation of the result by using the  Fermat principle. In the Fermat  approach it becomes transparent that in our slightly more general constellation the  deflection potential is  given by $\frac{1}{2}(-h_{00}+ h_{jj})$. The authors use a conformally stationary spacetime with metric  $ds^2=e^{2U}(dt-w_idx^i)^2 - e^{2U}dl^2$ (eq. 3.35). The transition from  (3.35) to the Fermat principle (3.39) $\delta \int_{\tilde{\gamma}}(w_idx^i + e^{-2U}dl)$ proceeds  via  specialization to  null curves, $ds^2 =0$, which leads to  $dt=w_idx^i + e^{-2U}dl$ (3.37). Generalizing 
 (3.35) to  $ds^2=e^{2U}(dt-w_idx^i)^2 - e^{2\tilde{U}}dl^2$ changes the null curve condition to (3.37') $dt=w_idx^i + e^{-(U+\tilde{U})}dl^2$ and the Fermat principle to
 \[ \delta \int_{\tilde{\gamma}} (w_idx^i + e^{-(U+\tilde{U})}dl)
 \] 
 (Similar at other places, e.g.,  \citep[chap. IX, \S3]{Synge:1960}.)
 For a static metric (with $w_i=0$) the spatial projections of light rays are geodesics of a Riemannian metric with arc length $e^{-(U+\tilde{U})}dl$, different from the ``physical'' arc length $e^{-\tilde{U}}dl$. A similar remark is given for the more special situation in \citep[p. 104]{Schneider/Ehlers/Falco}. 
\subsection{\small Comments on \citep{Scholz:MONDlike,Scholz:Clusters}  \label{Appendix Flaws}}
In \citep{Scholz:MONDlike} and \citep{Scholz:Clusters} it has been claimed that  a scalar field,  nonminimally coupled to the Hilbert action and with the  cubic kinetic term $L_{D\phi^3}$ (but without $L_{D^2\phi}$), can already bring new insight for the dark matter/modified gravity problem if is dealt with in framework of Weyl geometric gravity. 
Due to a flawed heuristic treatment of the weak field approximation, the energy momentum expression of the scalar field resulting from the variation of the non-minimal coupling (here eq. (\ref{eq Theta (H)})) was interpreted as an effective   contribution to the energy-momentum of the scalar field. But as we can  see from  (\ref{eq conformal Theta compensation}) the energy-momentum expression  (\ref{eq Theta (H)}) cancels with other terms of the Weyl geometric equation (most importantly the scale connection contribution to the Weyl geometric Einstein tensor). So this interpretation was wrong. On the other hand, some of the statements of the  papers, regarding the emulation of the deep MOND kinematics and  basic claims on the energy momentum of the scalar field, become correct in the present approach.\\[7em]


\addcontentsline{toc}{section}{\protect\numberline{}Bibliography}
\small
 \bibliographystyle{apsr}


\vspace*{2em}
{\bf Acknowledgements}: Many of the general reflections in this paper, and the motivation for writing it, are due to the common work in the project-group A3 ``LHC and gravity''  of the interdisciplinary group {\em Epistemology of the Large Hadron Collider} generously funded by the German {\em DFG} and the Austrian {\em FWF}. T. Scholz made me aware of the recent important work of Riess et al.; D. Lehmkuhl of the Geroch-Jang theorem. An anonymous referee encouraged me to discuss the bullet cluster and the question of the intermediate region between the effective Einstein and the modified gravity regime.

\end{document}